\documentclass[12pt,a4paper,reqno]{article}
\usepackage{amsmath,amsthm}
\usepackage{amsmath,amssymb}

\numberwithin{equation}{section}

\def\Dg {{\cal D}} 

\def\Hg {{\cal H}} 
\def\Xg {{\cal X}} 
\def\Yg {{\cal Y}} 
\def\R{{\bf R}}

\def\C{{\bf C}}

\def\Bb{{\bf B}}

\def\a{\alpha}
\def\b{\beta}

\def\d{\delta}

\def\ep{\varepsilon}

\def\th{\theta}

\def\m{\mu}

\def\r{\rho}
\def\s{\sigma}
\def\Si{\Sigma}

\newtheorem{theorem}{Theorem}[section]
\newtheorem{lemma}[theorem]{Lemma}
\newtheorem{proposition}[theorem]{Proposition}
\newtheorem{definition}[theorem]{Definition}
\newtheorem{corollary}[theorem]{Corollary}
\newtheorem{example}[theorem]{Example}
\newtheorem{assumption}[theorem]{Assumption}
\newtheorem{remark}[theorem]{Remark}
\newtheorem{problems}[section]{Problem}

\def\lbeq(#1){\label{eqn:#1}}
\def\refeq(#1){{\rm (\ref{eqn:#1})}}
\def\lbth(#1){\label{th:#1}}
\def\lbsec(#1){\label{sec:#1}}
\def\refsec(#1){{\rm Section \ref{sec:#1}}}
\def\refsecs(#1,#2){{\rm Sections \ref{sec:#1} and \ref{sec:#2}}}
\def\refth(#1){{\rm Theorem \ref{th:#1}}}
\def\refths(#1,#2){{\rm Theorems \ref{th:#1} and \ref{th:#2}}}
\def\refthb(#1){{\bf Theorem \ref{th:#1}}}
\def\lbprp(#1){\label{prp:#1}}
\def\refprp(#1){{\rm Proposition \ref{prp:#1}}}
\def\lblm(#1){\label{lm:#1}}
\def\reflm(#1){{\rm Lemma  \ref{lm:#1}}}
\def\reflms(#1,#2){{\rm Lemmas \ref{lm:#1} and \ref{lm:#2}}}
\def\lbcor(#1){\label{cor:#1}}
\def\refcor(#1){{\rm Corollary \ref{cor:#1}}}
\def\lbrm(#1){\label{rm:#1}}
\def\refrm(#1){{\rm Remark \ref{rm:#1}}}
\def\lbexm(#1){\label{exm:#1}}
\def\refexm(#1){{\rm Example \ref{rm:#1}}}
\def\lbass(#1){\label{ass:#1}}
\def\refass(#1){{\rm Assumption \ref{ass:#1}}}
\def\refasss(#1, #2){{\rm Assumptions \ref{ass:#1} and \ref{ass:#2}}} 
\def\lbdf(#1){\label{df:#1}}
\def\refdf(#1){{\rm Definition \ref{df:#1}}}

\def\bgdf{\begin{definition}}
\def\eddf{\end{definition}}
\def\bgth{\begin{theorem}}
\def\edth{\end{theorem}}
\def\bglm{\begin{lemma}}
\def\edlm{\end{lemma}}
\def\bgprp{\begin{proposition}}
\def\edprp{\end{proposition}}
\def\bgcor{\begin{corollary}}
\def\edcor{\end{corollary}}
\def\bgexm{\begin{example}}
\def\edexm{\end{example}}
\def\bgpf{\begin{proof}}
\def\edpf{\end{proof}}
\def\bgrm{\begin{remark}}
\def\edrm{\end{remark}}
\def\bgrms{\begin{remarks}}
\def\edrms{\end{remarks}}
\def\bgass{\begin{assumption}}
\def\edass{\end{assumption}}
\def\ben{\begin{enumerate}}
\def\een{\end{enumerate}}
\def\bgeq{\begin{equation}}
\def\edeq{\end{equation}}
\def\bgds{\begin{description}}
\def\edds{\end{description}}
\def\bgpbs{\begin{problems}}
\def\edpbs{\end{problems}}

\def\ben{\begin{enumerate}}
\def\een{\end{enumerate}}
\def\bqn{\begin{equation}}
\def\eqn{\end{equation}}
\def\bqna{\begin{eqnarray}}
\def\eqna{\end{eqnarray}}
\def\bgmul{\begin{multline*}}
\def\edmul{\end{multline*}}
\def\ep{\varepsilon}

\def\ph{\varphi}

\def\la{\langle}
\def\ra{\rangle}

\def\cos{{\rm cos}}

\def\ax{{\la x \ra}}

\def\br{\begin{array}}
\def\er{\end{array}}
\def\lap{\Delta}

\newcommand {\pa}{\partial}

\def\tL={{\tilde{L}}}
\def\loc{{\rm loc}}

\begin{document}
\allowdisplaybreaks

\title{Schr\"odinger equations 
with time-dependent strong magnetic fields}
\author{Daisuke Aiba and Kenji Yajima\footnote{
Department of Mathematics, Gakushuin University, 
1-5-1 Mejiro, Toshima-ku, Tokyo 171-8588, Japan.
Supported by JSPS grant in aid for scientific 
research No. 22340029 }
}
\date{}

\allowdisplaybreaks
\maketitle

\vspace{-0.5cm}
\begin{center}
{\it To the memory of late Professor Vladimir S. Buslaev}
\end{center}

\vspace{0.2cm}

\begin{abstract} 
We consider $d$-dimensional time dependent Schr\"odinger 
equations 
$i\pa_t u = H(t)u$, $H(t)=-(\pa_x-iA(t,x))^2+ V(t,x)$ 
in the Hilbert space $\Hg=L^2(\R^d)$ of square 
integrable functions. We assume $V(t,x)$ and $A(t,x)$ 
are almost critically singular with respect to spatial 
variables $x\in \R^d$ both locally and at infinity for 
the operator $H(t)$ to be essentially 
selfadjoint on $C_0^\infty(\R^d)$. In particular, when 
magnetic fields $B(t,x)$ 
produced by $A(t,x)$ are very strong at infinity, 
$V(t,x)$ can explode to the negative infinity like 
$-\th |B(t,x)|-C(|x|^2+1)$ for some $\th<1$ and $C>0$. 
We show that equations uniquely generate 
unitary propagators in $\Hg$ under suitable 
conditions on the size and singularities of time 
derivatives of potentials $\dot V(t,x)$ and 
$\dot A(t,x)$.   
\end{abstract}

\section{Introduction, Theorem} 
We consider time-dependent Schr\"odinger equations 
\bqn \lbeq(Seq)
i\pa_t u =  H(t) u(t) \equiv 
- \nabla_{A(t)}^2 u  + V(t,x) u, \quad 
\nabla_{A(t)} = \nabla - iA(t,x)
\eqn 
in the Hilbert space $\Hg= L^2(\R^d)$ of square 
integrable functions, where $A(t,x)=(A_1(t,x), \dots, A_d(t,x))\in \R^d$ and 
$V(t,x)\in \R$ are respectively magnetic vector 
and electric scalar potentials. We study the existence 
and the uniqueness of unitary propagators for Eqn. 
\refeq(Seq), 
continuing the previous work \cite{Yap} of the 
second author. 

In accordance with the requirement of quantum mechanics
we say that a function $u(t,x)$ of 
$(t,x) \in \R \times \R^d$ is a solution of \refeq(Seq) if 
it satisfies the following properties: 
\ben 
\item[{\rm (1)}] $u(t,\cdot)$ is a continuous function 
of $t\in \R$ with values in $\Hg$ and 
$\|u(t,\cdot)\|_{L^2}$ is independent of $t \in \R$. 
\item[{\rm (2)}] $u(t,x)$ satisfies Eqn. \refeq(Seq) in 
the sense of distributions. 
\een 

Suppose that there exists a dense subspace 
$\Si\subset \Hg$ such that, for every $s\in \R$ and 
$\ph \in \Si$, Eqn. \refeq(Seq) 
admits a unique solution $u(t,x)$ which satisfies the 
initial condition $u(s,x)=\ph(x)$ and that 
$u(t,\cdot) \in \Si$ for every $t\in \R$.    
Then the solution operator 
$\Si \ni \ph \mapsto u(t,\cdot)$ extends 
to a unitary operator $U(t,s)$ in $\Hg$ and the 
two parameter family of operators 
$\{U(t,s)\colon -\infty<t,s<\infty \}$ satisfies 
the following properties:   
\ben 
\item[{\rm (a)}]  $U(t,s)$ is unitary and 
$(t,s) \mapsto U(t,s)\in \Bb(\Hg)$ is 
strongly continuous.   
\item[{\rm (b)}] $U(t,s)U(s,r)=U(t,r)$ and 
$U(t,t) = {\bf 1}$ for every $-\infty<t,s,r<\infty$. 
\item[{\rm (c)}] $U(t,s)\Si = \Si$ and,  
for every $\ph \in \Si$, 
$u(t,x)= (U(t,s)\ph)(x)$ satisfies Eqn. \refeq(Seq) 
in the sense of distributions.
\een

\bgdf
We say a two parameter family of operators 
$\{U(t,s)\colon -\infty<t,s<\infty \}$ is a unitary 
propagator for \refeq(Seq) on a dense set $\Si$ if it 
satisfies properties {\rm (a), (b)} and {\rm (c)} above. 
\eddf 

Thus, the existence of a unique unitary propagator on 
a dense subspace of $\Hg$ implies that 
Schr\"odinger equation \refeq(Seq) generates a unique 
quantum dynamics on $\Hg$. When $A$ and $V$ are 
$t$-independent, it is well known 
that the existence of a unique unitary propagator 
on $\Hg$ is equivalent to the essential selfadjointness 
of Hamiltonian $-\nabla_{A}^2 + V$ on 
$C_0^\infty(\R^d)$. The problem of essential 
selfadjointness has long and extensively been studied by 
many authors and it has an extensive literature. 
We record here following two theorems, \refth(LS) of 
Leinfelder and Simader(\cite{LS}) and \refth(Iwatsuka) 
of Iwatsuka(\cite{Iwatsuka}) which are relevant to 
the present work. 
We need some notation: $(1+|x|^2)^{1/2}= \ax$; 
$L^p= L^p(\R^d)$, $1 \leq p \leq \infty$ are 
Lebesgue spaces and $L^p_{\loc}=L^p_{\loc}(\R^d)$ are 
their localizations; $\|u\|_p$ is the norm of 
$L^p$,  $\|u\|=\|u\|_2$ and $(u,v)$ is the 
inner product of $u,v \in \Hg$. A function $W(x)$ 
is said to be of Stummel class if it satisfies the 
property that 
\bqn \lbeq(stummel) 
\lim_{\ep\to 0} \sup_{x\in \R^d} 
\int_{|x-y|<\ep} 
\frac{|W(y)|^2}{|x-y|^{d-4}}dy =0,  
\eqn 
where $|x-y|^{4-d}$ should be replaced by $|\log |x-y||$ 
if $d=4$ and by $1$ if $1 \leq d \leq 3$.

\bgth \lbth(LS) Let $A\in L^4_{\loc}$ and  
$\nabla\cdot A \in L^2_{\loc}$. Let $V=V_1+ V_2$ 
with $V_1\in L^2_{\loc}$ and $V_2$ of Stummel class. 
Suppose that, 
for a constant $C_\ast>0$,  
\bqn \lbeq(low-con) 
V_1(x)\geq -C_\ast \ax^2.
\eqn 
Then, $H=-\nabla_A^2 + V$ is essentially 
selfadjoint on $C_0^\infty(\R^d)$. 
\edth 

It can be easily seen that conditions in \refth(LS) 
are also necessary as far as smoothness is 
concerned. However, condition \refeq(low-con) on 
on $V$ at infinity can be substantially relaxed if 
the magnetic field $B(x)= (B_{jk}(x))$ produced by $A$,, 
$B_{jk}= \pa_j A_k - \pa_k A_j$, $\pa_j = \pa/\pa x_j$,  
grows rapidly at infinity. We define 
\[
|B(x)|= \Big(\sum_{j<k} |B_{jk}(x)|^2\Big)^\frac12.
\]

\bgth \lbth(Iwatsuka) 
Let $\r(r)$ be a continuous function  of $r\geq 0$ 
such that   
\[
\int_0^\infty \r(r)^{-1}dr = \infty. 
\]
Suppose that $A$ and $V$ are $C^\infty$ and they  
satisfy that, for constants $C_\a$, 
\begin{gather} \lbeq(beta)
|\pa_x^\a B(x)|\leq C_\a \r(|x|)^{|\a|}(|B(x)|+1), \quad 
|\a|=1,2; \\
\lbeq(bbcon) 
|B(x)| + V(x) \geq -\r(|x|)^2 .
\end{gather}
Then, $H=-\nabla_A^2 + V$ is essentially selfadjoint 
on $C_0^\infty(\R^d)$.
\edth 
We remark that, by virtue of condition \refeq(beta), 
magnetic fields which behave too wildly at infinity, 
e.g. $|B(x)|\geq C\exp(\ax^{2+\ep})$ or 
$|B(x)| = C \cos(e^{\ax^{2+\ep}})$ for some $C>0$ 
and $\ep>0$, are {\it excluded} in \refth(Iwatsuka).  
To the best knowledge of authors, it is unknown 
whether or not \refth(Iwatsuka) remains true without 
this condition.  

We now state main results of this paper. 
We want to remark beforehand that, by virtue of 
assumptions on time derivatives, $A(t,x)$ and $V(t,x)$ in 
following theorems may be considered as perturbations 
of time frozen  potentials $A(t_0,x)$ and $V(t_0,x)$ 
respectively, $t_0$ being chosen arbitrarily. 

\bgdf $M(\R^d)$ is the space of real valued 
functions $Q(x)$ of class $C^1(\R^d)$ which satisfy 
for a positive constant $C>0$ that  
\bqn \lbeq(Q-qua)
Q(x) \geq C \ax \ \mbox{and}\ 
|\nabla Q(x)| \leq C \ax Q(x).
\eqn 
\eddf 

For $Q \in M(\R^d)$, $-\lap + Q(x)^2$ 
is essentially selfadjoint on $C_0^\infty(\R^d)$ 
(see \refth(LS)) and {\it hereafter $L_Q$ will denote 
its unique selfadjoint extension}. 
$L_Q\geq -\lap + C^2 x^2$ and $L_Q$ is positive definite;  
we have  
\begin{gather}
D(L_Q)=\{u \in \Hg \colon \lap u , \ Q\nabla u, \ 
Q^2 u \in \Hg\}, \\
C^{-1}\|L_Qu\| \leq \|\lap u\| + \|Q\nabla u\| + \|Q^2 u\| 
\leq C \|L_Qu\| , \quad u \in D(L_Q) \lbeq(dom-LQ)
\end{gather} 
for a constant $C>0$ (see the proof of \reflm(previous)). 

For Banach spaces 
$\Xg$ and $\Yg$, $\Bb(\Xg,\Yg)$ is the Banach space 
of bounded operators from $\Xg$ to $\Yg$ and 
$\Bb(\Xg)=\Bb(\Xg,\Xg)$. 
We say $f(t,x)$ is of class $C^\a(\R_x^d)$ 
if it is of class $C^\a$ with respect to variables 
$x \in \R^d$. Multiplication operators by $V(t,\cdot)$, 
$A(t,\cdot)$ and etc. are denoted by 
$V(t)$, $A(t)$ and etc. respectively;  
$\dot{A} (t,x)=\pa_t A(t,x)$ and 
$\dot{V}(t,x)=\pa_t V(t,x)$ are time derivatives.   
The letter $C$ denotes various constants whose exact 
values are not important and they may differ at each 
occurrence. 

First two theorems, \refths(Yap,Yap-f), 
may respectively be thought of as time dependent 
versions of \refth(LS) and its form version. 
$I$ is an interval. Under the assumption of \refth(Yap), 
operators $H_0(t)=-\nabla_{A(t)}^2+V(t,x)+C(t)\ax^2$ 
and $H(t)=-\nabla_{A(t)}^2 + V(t,x)$ 
are essentially selfadjoint on $C_0^\infty(\R^d)$ 
by virtue of \refth(LS).  
We denote their selfadjoint extensions again by 
$H_0(t)$ and $H(t)$. 

\bgth \lbth(Yap) 
Suppose $A$ and $V$ satisfy following conditions: 
\ben 
\item[{\rm (1)}] $A(t,\cdot) \in L^4_{\loc}$ and 
$\nabla_x\cdot A(t,\cdot) \in L^2_{\loc}$ for 
all $t \in I$. 
\item[{\rm (2)}] $V=V_1+ V_2$ with $V_1$ and $V_2$ 
such that $V_1(t,\cdot) \in L^2_{\loc}$ for 
$t \in I$ and $V_2(t,\cdot)$ of Stummel class 
uniformly for $t \in I$. 
There exist a continuous function $C(t)$ and  
$Q(x) \in M(\R^d)$ such that 
\bqn \lbeq(LBD-1)
V_1(t,x) + C(t)\ax^2 \geq Q(x)^2, \quad (t,x) 
\in I \times \R^d. 
\eqn 
\item[{\rm (3)}] For {\rm a.e.} $x \in \R^d$, 
$A(t,x)$ and $V(t,x)$ are absolutely continuous 
(AC for short in what follows) with 
respect to $t\in I$ and multiplication 
operators in $\Hg$ by following functions 
are all $L_Q$-bounded uniformly for 
$t\in I$: 
\[
\dot V(t,x), \quad \nabla_x\cdot \dot{A}(t,x), \quad  
\dot{A}(t,x)^2, \quad \pa_{x_j}\{(\dot{A}(t,x)^2)\},\ 
j=1,\dots, d. 
\]  
\een  
Then, following statements are satisfied: 
\ben 
\item[{\rm (a)}] 
$H_0(t)$ has $t$-independent domain $\Dg$ such that  
$\Dg \subset D(H(t))$. We equip $\Dg$ with the 
graph norm of $H_0(t_0)$, $t_0\in I$ being arbitrary. 
\item[{\rm (b)}] There uniquely exists 
a unitary propagator $\{U(t,s) \colon t, s \in I\}$ 
for \refeq(Seq) on $\Hg$ with following 
properties: $U(t,s) \in \Bb(\Dg)$; for $\ph \in \Dg$, 
$U(t,s)\ph$ is continuous in $\Si$ with respect to 
$(t,s)$,  of class $C^1$ in $\Hg$ and it satisfies 
\bqn \lbeq(dieq-1)
i\pa_t U(t,s)\ph = H(t)U(t,s) \ph, \quad 
i\pa_s U(t,s)\ph = -U(t,s) H(s) \ph. 
\eqn
\een 
\edth 

A remark on condition \refeq(LBD-1) which corresponds 
to \refeq(low-con) of \refth(LS) is in order since they 
look differently from each other. As was 
mentioned above we are considering Eqn. \refeq(Seq) 
when $A(t,x)$ and $V(t,x)$ satisfy conditions of 
\refth(LS) for every fixed $t \in \R$, 
in particular, that 
\bqn \lbeq(t-con) 
V_1 (t,x) \geq -C_\ast(t) \ax^2 
\eqn  
for a continuous $C_\ast(t)$. Then, if we choose 
$C(t)=C_\ast(t) + C$, $V_1(t,x)$ satisfies \refeq(LBD-1) 
with $Q(x)^2=C \ax^2 \in M(\R^d)$, $C$ being an 
arbitrarily large constant. However, this is 
the {\it worst} case conceivable and 
$V_1 (t,x)$ may rapidly grow to positive infinity 
as $|x|\to \infty$, in which case 
$V_1(t,x)$ certainly satisfies \refeq(t-con). If 
$V_1(t,x)$ increases the faster as $|x|\to \infty$, 
then $Q(x)$ of \refeq(LBD-1) may be taken the larger,  
condition (3) becomes the less restrictive and 
the class of potentials accommodated by the theorem 
becomes the wider. Condition \refeq(LBD-1) 
is formulated for studying these cases simultaneously. 
Similar remark applies to conditions \refeq(QFM), 
\refeq(bbcon-th-time) and \refeq(QFM-Iwap) in 
following theorems. 

When $V$ is spatially more singular than in \refth(Yap), 
we use quadratic form formalism. The following is  
a form version of \refth(Yap). A function 
$W(t,x)$ is said to be of 
Kato class uniformly for $t \in I$, if 
\bqn \lbeq(kato-type)
\lim_{\ep \to 0} \sup_{t\in I, x \in \R^d} 
\int_{|x-y|<\ep} \frac{|W(t,y)|}{|x-y|^{d-2}} dy =0, 
\eqn 
where $|x-y|^{2-d}$ should be replaced by $|\log |x-y||$ 
if $d=2$ and by $1$ if $d=1$. 
We write $q(u,u)=q(u)$ for quadratic forms $q(u,v)$.

\bgth \lbth(Yap-f) Suppose that $A$ and $V$ satisfy  
following conditions:
\ben 
\item[{\rm (1)}] $A(t,\cdot) \in L^2_{\loc}$ for 
every $t \in I$.
\item[{\rm (2)}] $V(t,x)=V_1(t,x) + V_2(t,x)$ with 
$V_1$ such that 
$V_1(t,\cdot) \in L^1_{\loc}(\R^d_x)$ for all $t\in I$ 
and $V_2(t,\cdot)$ of Kato class uniformly for 
$t \in I$. There exist a continuous function $C(t)$ 
and $Q\in M(\R^d)$ such that 
\bqn \lbeq(QFM)
V_1(t,x) + C(t) \ax^2 \geq Q(x)^2, \quad t \in I. 
\eqn 
\item[{\rm (3)}] $A$ and $V$ are AC 
with respect to $t$ for {\rm a.e.} $x\in \R^d$ and    
\bqn \lbeq(adot)
\|\dot A(t)L_Q^{-1/2}\|_{\Bb(L^2)} 
+ \| L_Q^{-1/2}\dot V(t) L_Q^{-1/2}\|_{\Bb(L^2)} 
\leq C, \quad t\in I 
\eqn 
for a constant $C>0$. 
\een
Then, following statements are satisfied:
\ben 
\item[{\rm (a)}] The quadratic form 
$q_0 (t)$ defined on $C_0^\infty(\R^d)$ by 
\[
q_0 (t)(u) = 
\int_{\R^d}(|\nabla_{A(t)} u |^2 + 
(V(t,x)+ C(t)\ax^2)|u|^2)dx 
\]
is strictly positive and closable; the closure 
$[q_0 (t)]$ has domain $\Yg$ independent 
of $t \in I$ and $\Yg \subset D(L_Q^\frac12)$. 
We equip $\Yg$ with the inner product  
$[q_0(t_0)](u,v)$ by choosing $t_0$ arbitrarily 
and denote by $\Xg$ its dual space with respect to 
the inner product of $\Hg$. We have 
$H(t)=-\nabla_{A(t)}^2 + V(t)\in \Bb(\Yg,\Xg)$ 
and $t \to H(t)\in \Bb(\Yg,\Xg)$ is norm continuous. 
\item[{\rm (b)}] There uniquely exists 
a unitary propagator for \refeq(Seq) on $\Yg$ with 
following properties: $U(t,s) \in \Bb(\Yg)$; 
for $\ph \in \Yg$, $U(t,s)\ph$ is continuous in $\Yg$ 
with respect to $(t,s)$, of class $C^1$ in $\Xg$ 
and satisfies equations \refeq(dieq-1). 
\een 
\edth 

Before stating time dependent versions of 
\refth(Iwatsuka), we generalize it for $V(x)$ which 
are locally as singular as those in \refth(LS) 
or in \refth(Yap-f) by slightly strengthening 
conditions \refeq(beta) and \refeq(bbcon) at infinity. 

\bgth \lbth(Iwa-sa) 
Let $A$ be of class $C^3$ and the magnetic 
field $B$ generated by $A$ satisfy for constants $C_\a$ 
that 
\bqn \lbeq(beta-a)
|\pa_x^\a B(x)|\leq C_\a \ax^{|\a|}(|B(x)|+1), \quad 
|\a|=1,2. 
\eqn 
Let 
$V(x)=V_1(x)+ V_2(x)$ with $V_1\in L^2_{\loc}$ and 
$V_2$ of Stummel class. Suppose that 
there exist constants $\th<1$ and $C_\ast>0$ such that  
\bqn \lbeq(bbcon-th) 
\th |B(x)| + 
V_1(x)\geq - C_\ast \ax^2, \quad x \in \R^d .   
\eqn  
Then, 
$L= -\nabla_A^2 + V$ is essentially selfadjoint 
on $C_0^\infty(\R^d)$ and the domain of 
its selfadjoint extension $H$ is given by   
$D(H) = \{ u \in \Hg \colon -\nabla_{A}^2 u + Vu \in \Hg\}$.\edth 

\bgth \lbth(Iwa-sa-f)
Let $A(x)$ and $B(x)$ be as in \refth(Iwa-sa). 
Let $V(x)=V_1(x) + V_2(x)$ with  
$V_1\in L^1_{\loc}(\R^d_x)$ and $V_2$ of Kato class. 
Suppose that there exist constants $\th<1$ and $C_\ast$ 
such that \refeq(bbcon-th) is satisfied.  Define 
\bqn \lbeq(tv1def)
\tilde V_1(x) = V_1(x)+ (C_\ast+C_1)\ax^2
\eqn
with a sufficiently large constant $C_1$. 
Then, following statements are satisfied: 
\ben 
\item[{\rm (1)}] The quadratic form $q_0$ on 
$C_0^\infty(\R^d)$ defined by 
\bqn \lbeq(q0q-s)
q_0(u) 
=\|\nabla_{A} u\|^2 + ((\tilde V_1+V_2)u, u) 
\eqn 
is bounded from below and closable. The closure has 
domain  
\bqn \lbeq(q0q-dom)
D([q_0])=\{ u \in L^2 \colon \nabla_{A} u \in L^2, 
\ \ (|B| + |\tilde V_1|+ \ax^2)^{1/2}u \in L^2\}.
\eqn 
For $u \in D([q_0])$, we have $V_2 |u|^2 \in L^1$   
and $[q_0](u)$ is given by \refeq(q0q-s). 
\item[{\rm (2)}] The selfadjoint operator $H_0$ defined 
by $[q_0]$ is given by 
\begin{gather}
H_0 u = -\nabla_{A}^2 u + (\tilde V_1+V_2) u, \\
D(H_0) = \{u \in D([q_0]), \ 
-\nabla_{A}^2 u + (\tilde V_1 + V_2 )u \in L^2\}.  
\end{gather}
\een
\edth 

Suppose that $A$ and $V$ satisfy conditions 
of \refth(Iwa-sa), then they also satisfy those 
of \refth(Iwa-sa-f), and the operator $H_0$ defined 
in \refth(Iwa-sa-f) is essentially selfadjoint on 
$C_0^\infty(\R^d)$ and $D(H_0)=\{u \in L^2 \colon 
(-\nabla^2_A + \tilde V_1 + V_2)u\in L^2\}$. This 
follows from the fact that selfadjoint operators 
admit no proper selfadjoint extensions.  

\refths(Iwap,Iwap-f) in what follows are time 
dependent versions of \refths(Iwa-sa,Iwa-sa-f) 
respectively. Under assumptions of \refth(Iwap)    
\[
H(t)= -\nabla_{A(t)}^2 + V(t,x)\ \ \mbox{and} \ \      
H_0(t)= -\nabla_{A(t)}^2 + V(t,x)+ (C(t)+C_1)\ax^2
\] 
are essentially selfadjoint on 
$C_0^\infty(\R^d)$ by virtue of \refth(Iwa-sa). 
We denote their selfadjoint extensions again by 
$H(t)$ and $H_0(t)$. 

\bgth \lbth(Iwap) 
Suppose that $A$ and $V$ satisfy following conditions:
\ben 
\item[{\rm (1)}] $A(t,x) \in C^3(\R^d_x)$ 
for all $t \in I$ 
and the magnetic field $B(t,x)$ generated by $A(t,x)$ 
satisfies, for constants $C_\a>0$, 
\bqn \lbeq(beta-ass)
|\pa_x^\a B(t,x)| \leq C_\a \ax^{|\a|} \la B(t,x)\ra, 
\ \ |\a|=1,2, \ \ (t,x) \in I \times \R^d.  
\eqn 
\item[{\rm (2)}] $V(t,x)=V_1(t,x)+ V_2(t,x)$ with 
$V_1 (t, \cdot) \in L^2_{\loc}(\R^d_x)$ 
for all $t \in I$ and $V_2(t,\cdot)$ of Stummel class 
uniformly with 
respect to $t \in I$. There exist a 
constant $\th<1$, a continuous function $C(t)$ and 
$Q \in M(\R^d)$ such that 
\bqn \lbeq(bbcon-th-time) 
\th |B(t,x)| + V_1 (t,x) + C(t)\ax^2 \geq Q(x)^2,
\quad (t,x) \in I \times \R^d.   
\eqn  
\item[{\rm (3)}] For {\rm a.e.} $x \in \R^d$, 
$A(t,x)$ and $V(t,x)$ are AC with respect to $t\in I$. 
Time derivatives satisfy, for a constant $C>0$, that 
\[
|\nabla_x\cdot \dot{A}(t,x)|+ |\dot{A}(t,x)|^2 
+ |\nabla_x (\dot{A}(t,x)^2)| \leq C Q(x)^2, \quad 
(t,x) \in I \times \R^d; 
\]
and that $\dot V(t,x)= W_0(t,x)+W_1(t,x)+ W_2(t,x)$ 
such that 
\[
\|Q^{-2+j}W_j(t)(-\lap +1)^{-j/2}\|_{\Bb(\Hg)}\leq C, 
\quad t\in I, \ \ j=0,1,2.  
\]
\een
Then, following statements are satisfied for a 
sufficiently large $C_1>0$: 
\ben 
\item[{\rm (a)}] Domain $\Dg$ of $H_0(t)$ is 
independent of $t\in I$ 
and $\Dg \subset D(H(t))$ for all $t \in I$. 
Equip $\Dg$ with the graph norm of $H_0(t_0)$, 
$t_0$ being arbitrarily. 
\item[{\rm (b)}] There uniquely exists a 
unitary propagator $\{U(t,s) \colon t, s \in I\}$ on 
$\Hg$ for \refeq(Seq) such that 
$U(t,s) \in \Bb(\Dg)$;   
for $\ph \in \Dg$, $U(t,s)\ph$ is continuous   
with respect to $(t,s)$ in $\Dg$, of class $C^1$ in $\Hg$ 
and satisfies \refeq(dieq-1).
\een
\edth

\bgth \lbth(Iwap-f)
Let $A(t,x)$ and $B(t,x)$ be as in \refth(Iwap). Suppose    \ben 
\item[{\rm (1)}] $V(t,x)=V_1(t,x) + V_2(t,x)$ with 
$V_1(t,\cdot) \in L^1_{\loc}(\R^d_x)$ for all $t \in I$ 
and $V_2(t,\cdot)$ of Kato class uniformly with respect 
to $t \in I$. There exist a $\th<1$, 
a continuous function $C(t)$ and $Q \in M(\R^d)$ such 
that     
\bqn \lbeq(QFM-Iwap)
\th |B(t,x)|+ 
V_1(t,x) + C(t) \ax^2 \geq Q(x)^2, 
\quad (t,x) \in I \times \R^d.
\eqn 
\item[{\rm (2)}] $V(t,x)$ is AC with 
respect to $t \in I$ for {\rm a.e.} $x\in \R^d$ and 
$\dot V(t,x)$ satisfies, for a constant $C>0$, 
\bqn \lbeq(previous)
\|L_Q^{-1/2}|\dot V(t)|L_Q^{-1/2}\|_{\Bb(L^2)} \leq C, 
\quad t \in I.
\eqn 
\een
Let $\tilde V= V + (C(t)+C_1)\ax^2$ and 
$\tilde V_1 = V_1+ (C(t)+C_1)\ax^2$ for a sufficiently 
large constant $C_1>0$. Then, 
following statements are satisfied.  
\ben 
\item[{\rm (a)}] The quadratic form $q_0(t)$ 
on $C_0^\infty(\R^d)$ defined by 
\bqn \lbeq(q0q)
q_0(t) (u) 
=\|\nabla_{A(t)} u\|^2 + (\tilde V(t,x)u, u) 
\eqn 
is bounded from below and closable. 
Domain $\Yg$ of its closure $[q_0(t)]$ is given 
by \refeq(q0q-dom) with obvious changes. $\Yg$ 
is independent of $t$ and satisfies 
$\Yg \subset D(L_Q^{\frac12})$.
We equip $\Yg$ with the inner product $[q_0(t_0)](u,v)$, 
$t_0 \in I$ being arbitrarily and denote by $\Xg$  
its dual space with respect to the inner product of $\Hg$. 
For $t \in I$, define operator 
$H(t)$ from $\Yg$ to $\Xg$ by 
\[
(H(t)u, v)= (\nabla_{A(t)}u, \nabla_{A(t)}v) 
+ (V(t,x)u, v), \quad u, v \in \Yg. 
\]
Then,  
$H(t) \in \Bb(\Yg,\Xg)$ and it is norm continuous with 
respect to $t\in I$. 
\item[{\rm (b)}] There uniquely exists a unitary 
propagator for \refeq(Seq) on $\Yg$ such that 
$U(t,s)\in \Bb(\Yg)$;  for $\ph \in \Yg$, 
$U(t,s)\ph$ is continuous 
with respect to $(t,s)$ in $\Yg$, 
of class $C^1$ in $\Xg$ and satisfies \refeq(dieq-1).
Moreover, $\{U(t,s)\}$ extends to a strongly continuous 
family of bounded operators in $\Xg$.  
\een 
\edth 

We emphasize that in all theorems above no conditions 
are imposed on the behavior at infinity of the positive 
part of $V$ in contrast to strong size restrictions 
on its negative part. 

For the reference on the problem, we refer to the 
introduction of \cite{Yap} and we shall jump into 
the proof of Theorems immediately. We shall not prove 
\refths(Yap,Yap-f) because they are proved in  
{\rm \cite{Yap}} for the case $Q(x)=C\ax$ 
and the proof goes through for the present cases with 
obvious changes, and because the proof of 
\refths(Iwap,Iwap-f) which we shall be devoted to in what 
follows basically patterns after that of \cite{Yap}, 
though several new estimates are necessary. 

The plan of paper is as follows. 
Section 2 collects some well known results which 
are necessary in subsequent sections. We prove 
selfadjointness theorems, \refths(Iwa-sa,Iwa-sa-f) 
in Section 3. In Section 4, we formulate and prove 
an estimate for the resolvent of 
$H_1(t)=-\nabla_{A(t)}^2 + V_1(t,x) + (C(t)+C_1)\ax^2$ 
which replaces the diamagnetic inequality 
(cf. \cite{CFKS}). 
We emphasize that it is hopeless to have 
standard diamagnetic inequality for this operator 
since the scalar potential 
$W(t,x)=V_1(t,x) + (C(t)+C_1)\ax^2$ of $H_1(t)$ 
can wildly diverge to negative infinity as 
$|x|\to \infty$ and $-\lap + W(t,x)$ is {\it not} in 
general essentially selfadjoint 
on $C_0^\infty(\R^d)$. We prove \refths(Iwap,Iwap-f) 
in Sections 5 and 6 respectively by 
using materials prepared in preceding sections.    

\section{Preliminaries} 
In this section, we recall Kato's abstract theory of 
evolution equations which the proof of Theorems will  
eventually relies upon, and Iwatsuka's identity which 
will be used for deriving various estimates necessary 
for applying Kato's theory.   

\subsection{Kato's abstract theory for evolution equations}

As in the previous paper \cite{Yap}, 
\refths(Iwap,Iwap-f) will be proven by 
applying the following abstract theorem. The 
theorem is the consequence of Theorem 5.2, Remarks 5.3 
and 5.4 of Kato's seminal paper \cite{Ka-evo1}. 

\bgth \lbth(se) Let $\Xg$ and $\Yg$ be a pair of 
Hilbert spaces such that $\Yg \subset \Xg$ continuously 
and densely. Let $\{A(t), t \in I\}$, 
$I$ being an interval, be a family 
of closed operators in $\Xg$ with dense domain $D(A(t))$ 
such that $\Yg \subset D(A(t))$ for every $t \in I$ 
and $I \ni t \to A(t) \in \Bb(\Yg, \Xg)$ is norm 
continuous. 
Suppose that following conditions are satisfied:
\ben 
\item[{\rm (1)}] For every $t\in I $, 
there exist inner products $(\cdot, \cdot)_{\Xg_t}$ 
and $(\cdot, \cdot)_{\Yg_t}$ of $\Xg$ and $\Yg$ 
respectively 
which define norms equivalent to the original ones and 
which satisfy, for a constant $c>0$, 
\bqn \lbeq(xy-equiv) 
\|u\|_{\Yg_t}/\|u\|_{\Yg_s} \leq e^{c|t-s|}, \quad 
\|u\|_{\Xg_t}/\|u\|_{\Xg_s}\leq e^{c|t-s|}, \quad u \not=0.
\eqn  
\item[{\rm (2)}] If we let $\Xg_t$ and $\Yg_t$ be 
Hilbert spaces $\Xg$ and $\Yg$ with these inner products,  
$A(t)$ is selfadjoint in $\Xg_t$ and 
the part $\tilde A(t)$ of $A(t)$ in $\Yg_t$ 
is also selfadjoint in $\Yg_t$. 
\een
Then, there  uniquely exists a strongly continuous 
family of bounded operators $\{U(t,s) \colon t,s \in I\}$ 
in $\Xg$ that satisfies 
\ben 
\item[{\rm (a)}] $U(t,r)=U(t,s)U(s,r)$, 
$U(s,s)=I$ for every $t,s$ and $r \in I$. 
\item[{\rm (b)}] $U(t,s)\in \Bb(\Yg)$;   
for $\ph \in \Yg$, $U(t,s)\ph$ is continuous with respect 
to $(t,s)$ in $\Yg$, of class $C^1$ in $\Xg$ and it 
satisfies  
\bqn \lbeq(pa-de-ev)
\pa_ t U(t,s)\ph = -iA(t) U(t,s)\ph, \quad 
\pa_s U(t,s)\ph = i U(t,s)A(s)\ph.
\eqn 
\een
\end{theorem}

\subsection{Iwatsuka's Identity}
In \cite{Iwatsuka}, Iwatsuka has found an ingenious 
formula which rewrites Schr\"odinger operator 
$H=-\nabla_A^2 + V$ in the form of elliptic operators 
in which the magnetic field $B_{jk}=\pa_j A_k-\pa_k A_j$ 
appears explicitly, which he has used for proving 
\refth(Iwatsuka). We recall it here as we shall use it 
several times for deriving 
various estimates.  For the proof of following lemmas 
we refer to Iwatsuka's paper \cite{Iwatsuka}, 
formula (2.12) and proofs of Theorem 1.1 and Theorem 2.1 
therein. 
We denote $b\cdot a= {}^t b a$ for a vector 
$b$ and a matrix $a$. 

\bglm \lblm(Iwa-1)
Let $G(x)=\{G_{jk}\}$ be Hermitian matrix valued 
function and 
\[
G_{jk}= \a_{jk}+ i\b_{jk}, \ \mbox{for real valued}\ 
\a_{jk}=\a_{kj} \ \mbox{and}\  
\b_{jk}= - \b_{kj}, \ j,k=1, \dots,d;   
\]
$F(x)=\{F_j\}$ be complex vector field such that 
with real $A$ and complex $b$ 
\bqn \lbeq(fdef)
F(x)= A(x) + b(x) 
\eqn 
and $B(x)=\{B_{jk}\}$, $B_{jk}=\pa_j A_k - \pa_k A_j$. 
Then, we have the following identity: 
\begin{align}
-\nabla_{\overline{F}}\cdot G \nabla_F  
= -\nabla_A \cdot \a \nabla_A & + 
i\{2\Re(\overline{b}\cdot G) - (\nabla\cdot \b) \}\nabla_A 
\notag \\
& - \sum_{j<k}\b_{jk}B_{jk}  
+i \nabla \cdot (G b)+ \overline{b}\cdot G b. 
\lbeq(pot)
\end{align}
In particular, if $\a_{jk}=\d_{jk}$, Kronecker's delta 
and 
\bqn \lbeq(ab)
G_{jk} = \d_{jk}+i\b_{jk} \quad 
\mbox{and} \ \ b = \tfrac12 \nabla \cdot \b 
\eqn  
for a real valued skew-symmetric matrix $\{\b_{jk}\}$, 
then 
\begin{gather}
- \nabla_A^2 
= -\nabla_{\overline{F}}\cdot G \nabla_F + 
\sum_{j<k}\b_{jk} B_{jk}+ R,   
\lbeq(T=H) \\ 
R= \tfrac12 \sum_{j,k}\b_{jk} \pa_{j}b_{k}  
+ \tfrac14 b^2 . \lbeq(R)
\end{gather}
\edlm 

Real skew-symmetric $\b$ in \refeq(ab) is 
completely arbitrary for identity \refeq(T=H) 
and Iwatsuka's choice in \cite{Iwatsuka} is as follows:  
Take $\chi \in C^\infty([0,\infty))$ such that 
$\mbox{$\chi(r)=1$ for $0\leq r \leq 1/2$}, \   
\mbox{$\chi(r)=r^{-1}$ for $ r\geq 1$ }$ and 
\[
\mbox{$0<r\chi(r)\leq 1$ for all $r>0$}
\]
and define  
\bqn \lbeq(beta-def) 
\b(x) = \chi(|B(x)|)B(x). 
\eqn 
{\it In what follows, $\b(x)$ always denotes the function 
defined by \refeq(beta-def) and $b(x)$ and $R(x)$ 
are respectively defined by \refeq(ab) and \refeq(R) by 
using this $\b(x)$}. We write 
\[
|\pa B|= \sum_{|\a|=1,j<k} |\pa^\a B_{jk}| \quad 
\mbox{and} \quad 
|\pa^2 B|= \sum_{|\a|=2,j<k} |\pa^\a B_{jk}|.
\]

\bglm Suppose $A(x)$ and $B(x)$ satisfy \refeq(beta-a). 
Then: 
\begin{gather} |\b(x)|\leq 1, \quad 
\sum_{j<k}\b_{jk} B_{jk}= \chi(|B|)|B|^2 \geq |B| - 1, 
\lbeq(bB)\\
|\pa_x^\a \b| \leq C \ax^{|\a|}, \quad |\a|=1,2; 
\quad 
|b| \leq C \ax, \quad |R|\leq C \ax^2. \lbeq(bound)
\end{gather}
\edlm 

For real skew-symmetric $\tilde{\b}=(\tilde{\b}_{jk})$, 
we have 
(Proposition 4.1 of \cite{Iwatsuka}) that 
\bqn \lbeq(betabound)
-|\tilde{\b}| \leq i\tilde{\b} \leq |\tilde{\b}|, 
\quad |\tilde{\b}|
=\Big(\sum_{j<k} \tilde{\b}_{jk}^2 \Big)^\frac12 
\eqn 
in the sense of quadratic forms on $\C^d$. 
In what follows we shall use identity 
\refeq(T=H) by modifying $\b(x)$ of \refeq(beta-def) 
in various ways.

\section{Selfadjointness} \lbsec(self) 

We prove \refths(Iwa-sa,Iwa-sa-f) in this section. 
We take and fix $\ph\in C_0^\infty(\R^d)$ such that 
$0 \leq \ph(x) \leq 1$ for all $x\in \R^d$, 
\bqn \lbeq(ph)
\mbox{$\ph(x)=1$ for $|x|\leq 1$ and 
$\ph(x)=0$ for $|x|\geq 2$}.  
\eqn 
We set $\ph_n(x) = \ph(x/n)$ for $n=1,2, \dots$ 
and define for $0<\th\leq 1$  
\bqn \lbeq(bnth)
\b_{n,\th}(x)= \th \ph_n(x)\b(x).
\eqn 
The following lemma is obvious by virtue of 
\refeq(betabound). 

\bglm If we change $\b$ by 
$\b_{n,\th}(x)$, then \refeq(T=H) remains to hold with 
$G$, $b$ and $R$ being replaced by corresponding 
$G_{n,\th}$, $b_{n,\th}$, $R_{n,\th}$.
Matrix $G_{n,\th}$ satisfies  
\bqn \lbeq(lower) 
G_{n,\th}(x)= {\bf 1} + i\th\ph_n(x)\b(x) 
\geq {\bf 1}-\th, 
\quad x \in \R^d; 
\eqn 
and $b_{n,\th}$ and $R_{n,\th}$ satisfy 
corresponding estimates in \refeq(bound) 
uniformly with respect to $\th$ and $n$. 
\edlm 

\paragraph{Proof of \refthb(Iwa-sa)} 
The following is a modification of Kato's argument 
(\cite{Ka-sa1}). It suffices to show that the 
image of $L\pm i$, $R(L\pm i)$, is dense in 
$\Hg$. Thus we suppose that 
$f \in \Hg$ satisfies $f \perp R(L\pm i)$ 
and show $f=0$ then. We prove the $+$ case 
only. The proof for the other case is similar. 

We first assume $V_2=0$. Define, for $n=1,2, \dots$, 
$V_n(x) = \chi_{B_{2n}(0)}(x) V(x)$, where 
$B_{2n}(0)=\{x\in \R^d \colon |x|<2n\}$ and $\chi_F$ 
is the characteristic function of the set $F$, and 
\[
L_n = -\nabla_A^2 + V_n, \quad D(L_n)= C_0^\infty(\R^d).
\]
Since $V_n(x)$ is bounded from below, $L_n$ is essentially 
selfadjoint by virtue of \refth(LS). It follows that 
there exists 
$u_n \in C_0^\infty(\R^d)$ such that 
\bqn \lbeq(appro)
\|(L_n + i)u_n - f \| \leq 1/n , \quad n=1,2, \dots .
\eqn 
Then, $\|(L_n+i)u_n\|\leq \|f\|+ 1/n$ and 
\bqn \lbeq(unb)
\|u_n\| \leq \|(L_n+i)u_n\| \leq C, \quad 
\|L_n u_n \|\leq \|f\|+ \|u_n\|+1/n \leq C. 
\eqn 
Let $\ph_n(x)$ be as above. Then, 
$\ph_n(x)V_n(x) = \ph_n(x) V(x)$ and 
\[
\ph_n(x) (L_n + i)u_n = (L + i) \ph_n u_n 
+ 2(\nabla \ph_n) \nabla_A u_n + (\lap \ph_n) u_n. 
\]
It follows from \refeq(appro) that 
\begin{align}
\|f\|^2 
& = \lim_{n \to \infty}(\ph_n f, (L_n+i)u_n) 
= \lim_{n \to \infty}(f, \ph_n (L_n+i)u_n) \notag \\ 
& = \lim_{n \to \infty}
\{(f, (L+i)\ph_n u_n) 
+ 2(f,(\nabla \ph_n) \nabla_A u_n) + 
(f, (\lap \ph_n) u_n)\}. \lbeq(prodct) 
\end{align}
The first term on the right vanishes 
by the assumption and the third satisfies 
\[
|(f, (\lap \ph_n) u_n)|
\leq n^{-2} \|\lap \ph\|_\infty \|f\|\|u_n\| \to 0 
\quad (n\to \infty). 
\]
For estimating $\|\nabla_A u_n\|$, we use Iwatsuka's 
identity \refeq(T=H) with $\b_{2n,\th}$ defined by 
\refeq(bnth) with $2n$ replacing $n$, which produces 
\begin{gather} \lbeq(Lnre)
L_n= -\nabla_A^2 + V_n = -\nabla_{\overline{F_{2n,\th}}}
G_{2n,\th}\nabla_{F_{2n,\th}} + W_{2n,\th}, \\
F_{2n,\th}= A+ b_{2n,\th}, \quad 
W_{2n,\th}= V_n + \sum \b_{2n,\th,jk}B_{jk}+R_{2n,\th}. 
\end{gather}
Here $W_{2n,\th}$  satisfies, with a constant $C$ 
independent of $n$, that  
\bqn \lbeq(Wnth)
W_{2n,\th}(x) \geq -C n^2 , \quad n=1,2, \dots, 
\quad x \in \R^d. 
\eqn 
Indeed, for $|x|\leq 2n$, we have $\ph_{2n}(x)=1$ 
and \refeq(bbcon-th), \refeq(bB) and \refeq(bound) imply
\begin{align*}
W_{2n,\th} & = V + \th \sum \b_{jk}B_{jk}(x)+ \th^2 R \\
& \geq V + \th(|B| -1)+ \th^2 R 
\geq -C \ax^2 \geq -C n^2; 
\end{align*}
for $2n<|x|\leq 4n$, we have 
$V_n(x)=0$ and 
\[
W_{2n,\th} = \th \ph_{2n}(x)\sum \b_{jk}B_{jk}(x)+ 
R_{2n,\th}(x) 
\geq R_{2n,\th}(x) \geq -C n^2; 
\]
and, for $|x|\geq 4n$, $W_{2n,\th}(x)=0$. It follows 
by virtue of \refeq(lower) and \refeq(Lnre) that    
\begin{multline}
(1-\th)\|\nabla_{F_{2n,\th}} u_n\|^2 \leq 
(G_{2n,\th} \nabla_{F_{2n,\th}} u_n, 
\nabla_{F_{2n,\th}}u_n)\\  
= ((L_n - W_{2n,\th})u_n, u_n) 
\leq (L_n u_n, u_n) + Cn^2 \|u_n\|^2 
\leq C n^2.   \lbeq(cn2)
\end{multline}
Since $|b_{2n,\th}(x)|\leq C n$ by \refeq(bound), 
we then have  
\bqn \lbeq(AF)
\|\nabla_A u_n \|
\leq \|\nabla_{F_{2n,\th}} u_n \| + 
\|b_{2n,\th} u_n \| 
\leq \|\nabla_{F_{2n,\th}} u_n \| + C n \|u_n\| \leq Cn 
\eqn 
and $\|(\nabla \ph_n)\nabla_A u_n\|\leq 
n^{-1}\|\nabla\ph\|_\infty\|\nabla_A u_n \| \leq C$.  
It follows, since $\nabla\ph_{n}=0$ for $|x|\leq n$, that 
\[
|(f, (\nabla \ph_{n})\nabla_A u_n)|\leq 
C\|f\|_{L^2(|x|\geq n)}\to 0 
\]
as $n \to \infty$. Thus, the right of \refeq(prodct) 
vanishes and $f=0$ and $L$ is essentially  
selfadjoint on $C_0^\infty(\R^d)$.  

If $V_2\not=0$, we repeat the argument above,  
setting $V_n= \chi_{|x|\leq 2n} V_1+ V_2$. 
Since $V_2$ is of Stummel class, $L_n$ with this $V_n$ 
is essentially selfadjoint on $C_0^\infty(\R^d)$ by 
virtue of \refth(LS) and it suffices to show 
$(f,(\nabla \ph_n) \nabla_A u_n) \to 0$ as $n\to \infty$ 
for $u_n \in C_0^\infty(\R^d)$ of \refeq(appro). 
We use identity \refeq(Lnre) and obtain 
\[
(1-\th)\|\nabla_{F_{2n,\th}}u_n \|^2 
\leq (L_n u_n, u_n) -(V_2u_n, u_n)
+ Cn^2 \|u_n\|^2. 
\]
as in \refeq(cn2). This with \refeq(unb) implies as 
in \refeq(AF) that 
\[
\|\nabla_A u_n\|^2 \leq C(n^2 + |(V_2u_n, u_n)|).
\] 
Since $V_2$ is $-\lap$-form bounded with bound $0$,  
we have, for any $\ep>0$,  
\[
|(|V_2|u, u)| \leq \ep 
\|\nabla |u|\|^2 + C_\ep \|u\|^2 
\leq \ep \|\nabla_A u\|^2 + C_\ep \|u\|^2, \quad 
u \in C_0^\infty(\R^d).  
\]
It follows that $\|\nabla_A u_n \| \leq C n$ and 
$\lim_{n\to \infty}(f,(\nabla \ph_n) \nabla_A u_n)=0$ as 
previously. Thus, $L$ is essentially selfadjoint 
when $V_2 \not=0$ as well. The closure of $L$ is given 
by $H=L^\ast$ and it is standard that 
$D(L^\ast)
= \{u\in\Hg \colon -\nabla_A^2 u + V u \in L^2\}$ 
and this completes the proof. \qed 

\paragraph{Proof of \refthb(Iwa-sa-f)} 
We let $\th$ and $\tilde V_1$ be as in the theorem. 
Define  
\[
G_{\th_0} = {\bf 1} + i{\th_0} \b, \quad  
F_{\th_0}= A + {\th_0} b \quad 
\mbox{for $\th \leq \th_0 \leq 1$} 
\]
by replacing $\b$ and $b$ by $\th_0\b$ and $\th_0 b$ 
in \refeq(ab) and \refeq(fdef) respectively. 
We have  
\bqn \lbeq(iwa-theta)
-\nabla_{A}^2 + \tilde V_1 
= -\nabla_{\overline{F_{\th_0 }}}
G_{\th_0 } \nabla_{F_{\th_0 }} 
+ \tilde W_{{\th_0 }}, \quad 
\tilde W_{{\th_0 }}
= \tilde V_1+ {\th_0 } 
\sum \b_{jk}B_{jk}+{\th_0 }^2 R. 
\eqn 
We take the constant $C_1\geq 10$ large enough in 
the definition \refeq(tv1def) of $\tilde V_1$ 
so that $|R(x)|\leq 10^{-2} C_1 \ax^2$ and 
\bqn \lbeq(atode)
\tilde W_{\th} \geq \tilde V_1 + \th(|B|-1) + 
\th^2 R 
\geq C_1 \ax^2 -1 - |R| \geq 
\tfrac23 C_1 \ax^2 + 2 |R|.
\eqn 
We show that, for $\th<\th_0\leq 1$, 
there exist a $\th_0$-dependent constant $C_{\th_0}>0$ 
and a $\th_0$-independent $C>0$ such that 
\bqn \lbeq(W-rtveq)
C_{\th_0} (|B(x)|+ |\tilde V_1(x)|)+ \tfrac{C_1}{2}\ax^2 
\leq \tilde W_{\th_0} (x) 
\leq (|B(x)|+ |\tilde V_1(x)|+ C \ax^2). 
\eqn 
Indeed, the second inequality is obvious from 
\refeq(bound). The first is also evident if 
$\tilde V_1> 0$, since then  
$\tilde V_1 +\th |B| \geq C_1 \ax^2$ and 
\[
\tilde W_{\th_0} 
\geq \tilde V_1 + \th_0(|B|-1) + \th_0^2 R  
\geq \frac12 (|\tilde V_1| + \th_0 |B|+ C_1 \ax^2).
\]
To see the first for the case $\tilde V_1(x)<0$, 
we first estimate  
\begin{align*}
\tilde W_{\th_0} & =\tilde W_{\th} 
+ (\th_0-\th)\sum\b_{jk}B_{jk} + (\th_0^2-\th^2)R \notag \\
& \geq \tfrac23 C_1 \ax^2 + (\th_0-\th)(|B|-1) -|R| 
\geq \tfrac12 C_1 \ax^2+ (\th_0-\th)|B| 
\end{align*}
which holds irrespectively of the sign of $\tilde V_1$. 
If $\tilde V_1(x)<0$ we also have      
\begin{align*}
\tilde W_{\th_0}& = \tfrac{\th_0}{\th}\tilde W_{\th}+ 
\left(\tfrac{\th_0}{\th}-1\right)|\tilde V_1|+ 
\th_0(\th_0-\th)R \\
& \geq \tfrac{\th_0}{\th}(\tfrac23 C_1\ax^2+ 2|R|)+ 
\left(\tfrac{\th_0}{\th}-1\right)|\tilde V_1| -|R| 
\geq \tfrac23C_1 \ax^2 +  
\left(\tfrac{\th_0}{\th}-1\right)|\tilde V_1|. 
\end{align*}
Adding both sides of last two estimates and 
dividing by $2$, we obtain the first inequality of 
\refeq(W-rtveq) for the case $\tilde V_1(x)<0$. 

We define the quadratic 
form $q_1(u,v)$ for $u,v \in C_0^\infty(\R^d)$ by 
\bqn q_1(u,v) 
=(\nabla_A u, \nabla_A v) + (\tilde V_1 u, v). 
\lbeq(q1q) 
\eqn 
We have by virtue of 
Iwatsuka's identity \refeq(iwa-theta) for 
$\th_0$ replacing $\th$ that  
\bqn 
q_1(u,v)
= (G_{\th_0}\nabla_{F_{\th_0}} u, \nabla_{F_{\th_0}} v) + 
(\tilde W_{\th_0} u, v). 
\eqn 
Estimates $1-\th_0 \leq G_{\th_0} \leq 1+ \th_0$ 
and  \refeq(W-rtveq) imply for a constant $C>1$ that 
\begin{gather*}
(1-\th_0)\|\nabla_{F_{\th_0}}u\|^2 \leq 
(G_{\th_0}\nabla_{F_{\th_0}} u, \nabla_{F_{\th_0}}u)
\leq (1+\th_0)\|\nabla_{F_{\th_0}}u\|^2, \notag \\
C^{-1} \|(|B|+|\tilde V_1|+\ax^2)^{\frac12}u\|^2 
\leq (\tilde W_{\th_0} u, u) 
\leq C \|(|B|+|\tilde V_1|+\ax^2)^{\frac12}u\|^2. 
\end{gather*} 
It follows that quadratic forms 
$(G_{\th_0}\nabla_{F_{\th_0}} u, \nabla_{F_{\th_0}}v)$ 
and $(\tilde W_{\th_0} u, v)$ on $C_0^\infty(\R^d)$ 
are both closable and positive definite and 
their closures have respective domains 
$\{u \colon \nabla_{F_{\th_0}}u \in L^2\}$ and  
$\{u \colon (|B|+|\tilde V_1|+\ax^2)^\frac12 u \in L^2\}$. 
Thus, $q_1$ is closable, the closure $[q_1]$ 
has domain 
\begin{align}
D([q_1])& 
= \{u \in L^2 \colon \nabla_{F_{\th_0}} u \in L^2, \ 
(|B|+|\tilde V_1|+\ax^2)^\frac12 u \in L^2\} \lbeq(dmq1) \\
& =\{u \in L^2 \colon \nabla_{A} u \in L^2, 
\ (|B|+|\tilde V_1|+\ax^2)^\frac12 u \in L^2\} 
\lbeq(dmq1a)
\end{align}
and $[q_1](u)$ is given again by \refeq(q1q). 
Moreover, by making $C_1$ larger if necessary, 
we have from the first inequality of 
\refeq(W-rtveq) and that $|b_{\th_0}|\leq C \ax$ 
that 
\bqn \lbeq(q1-estimate)
[q_1](u) \geq (1-\th_0)\|\nabla_{A} u\|^2 
+ C\|(|B|+|\tilde V_1|+\ax^2)^{\frac12} u\|^2, \quad 
u \in D([q_1]).
\eqn  
We have $q_0(u,v)=q_1(u,v)+ (V_2 u,v)$. 
Since $V_2$ is of Kato-class, $V_2$ is $-\lap$-form 
bounded with bound $0$ and we have, for any $\ep>0$, 
\bqn \lbeq(kato-q)
(|V_2|u,u)\leq \ep \|\nabla_A u\|^2 + C_\ep\|u\|^2 
\eqn 
as in the proof of \refth(Iwa-sa). Hence the form 
$(|V_2|u,u)$ is $[q_1]$-bounded with bound $0$ and 
statements (1) and (2) of the theorem follow. 

We prove statement (3). 
We write $\tilde V= \tilde V_1+ V_2$. 
Let $u \in D(H_0)$. Then, $u \in D([q_0])$ 
and $\ax u$, 
$|\tilde V_1|^\frac12  u, |V_2|^\frac12 u \in \Hg$ 
and $\nabla_A u \in \Hg$. Hence,  
$\tilde V u \in L^1_{\loc}$ and $\nabla_A^2 u $ is well 
defined as distributions. It follows for any 
$v\in C_0^\infty(\R^d)$ that  
\[
(H_0 u, v) = [q_0](u,v)=
(\nabla_A u, \nabla_A v) + (\tilde V u, v)
=(-\nabla_A^2 u + \tilde V u, v).
\]
Hence $-\nabla_A^2 u + \tilde V u\in L^2$ and 
$H_0 u= -\nabla_A^2 u + \tilde V u$. 
Suppose on the contrary that $u \in D([q_0])$ satisfies  
$-\nabla_A^2 u + \tilde V u\in L^2$.  
Then, for any $v \in C_0^\infty(\R^d)$,  
\[
(-\nabla_A^2 u + \tilde V u, v)=[q_0](u,v) 
= (G_{\th_0} \nabla_{F_{\th_0}} u, \nabla_{F_{\th_0}} v) 
+ ((\tilde W_{\th_0}+V_2) u,v) 
\]
and this extends to all $v \in D([q_0])$ by virtue of 
the argument in the first part. Thus, 
$u \in D(H_0)$ and $H_0 u = -\nabla_A^2 u + \tilde V u$. 
This completes the proof.   \qed 

The following is a corollary of 
the proof of \refth(Iwa-sa-f). 
\bgcor Let conditions of \refth(Iwa-sa-f) be satisfied. 
Let $C_1$ be sufficiently large. Then, for a constant 
$C>0$, we have 
\bqn \lbeq(q1-est0)
\|\nabla_A u\|^2 + 
\|(|B|+|\tilde V_1|+\ax^2)^\frac12 u\|^2 \leq C [q_0](u) , 
\quad u \in D([q_0])
\eqn 
\edcor 

\section{Diamagnetic inequality} \lbsec(diam) 

In this section we assume that $A$ and $V$ satisfy 
the following conditions: 
\ben 
\item[{\rm (1)}] $A(x)\in C^3(\R^d)$ and  
$B(x)$ satisfies estimates \refeq(beta-a). 
\item[{\rm (2)}] $V=V_1+ V_2$ with $V_1\in L^1_{\loc}$ 
and $V_2$ of Kato class. 
\item[{\rm (3)}] There exists constants $0<\th<1$, 
$C_\ast>1$ and $Q\in M(\R^d)$ such that 
\bqn 
\th |B(x)|+ V_1(x) + C_\ast \ax^2 \geq Q(x)^2. 
\lbeq(QFM-s-imp) 
\eqn 
\een
We then define $q_0(u)$ and $q_1(u)$ respectively 
by \refeq(q0q-s) and \refeq(q1q) 
with $\tilde V_1(x)=V_1(x)+ (C_\ast+C_1)\ax^2$  
with sufficiently 
large constant $C_1$ such that results in the 
previous section are satisfied. We let $H_0$ and $H_1$ 
be selfadjoint operators defined by $[q_0]$ and $[q_1]$ 
respectively.  

\bglm \lblm(previous) Let $\th<\th_0<1$. 
There exists $C_{\th_0}>0$ 
such that for $C_1\geq C_{\th_0}$, we have 
the following estimate:
\begin{multline} \lbeq(apri-2)
(1-\th_0)\|\nabla_{F_{\th_0}} u\|^2 
+ \|Q^2 u\|^2 \\ + 
2(\th_0-\th)\|Q |B|^\frac12 u\|^2 
+C_1 \|\ax Q u\|^2 \leq \|H_1 u\|^2, \quad u \in D(H_1).
\end{multline}
\edlm 
\bgpf 
We use the notation of the proof of \refth(Iwa-sa-f). 
We have as in there   
\bqn \lbeq(below)
\tilde W_{\th_0} \geq Q(x)^2 + (\th_0-\th)|B(x)| + 
\tfrac23 C_1 \ax^2 
\eqn 
Let $u \in D(H_1)$. 
Then, $\nabla_{F_{\th_0}} u$, 
$\nabla_A u$, $\tilde W_{\th_0}^{1/2} u$ and $Qu$ 
all belong to $L^2(\R^d)$ by virtue  of \refeq(below) 
and, for $v \in C_0^\infty(\R^d)$, we have 
\begin{multline}\lbeq(11)
(G_{\th_0} Q \nabla_{F_{\th_0}} u, 
Q \nabla_{F_{\th_0} }v)= 
-(\nabla_{\overline{F_{\th_0}}}G_{\th_0} 
\nabla_{F_{\th_0}}u, Q^2 v) 
- (G_{\th_0} \nabla_{F_{\th_0}} u, \nabla(Q^2) v) \\
= (H_1 u, Q^2 v) -(\tilde W_{\th_0} u , Q^2 v) 
- (G_{\th_0} \nabla_{F_{\th_0}} u, \nabla(Q^2) v). 
\end{multline}
Using $\ph_n(x)$ of the proof of \refth(Iwa-sa) and 
Friedrich's mollifier $j_\ep$, we define 
$v_{\ep,n}=j_\ep \ast (\ph_n^2 u)$ for $0<\ep<1$ and 
$n=1,2, \dots$. Then, 
$v_{\ep,n}\in C^\infty_0(\R^d)$, is supported by 
the ball $B_{2(n+1)}(0)$ 
and $v_{\ep,n} \to \ph_n^2 u$ in the 
Sobolev space $H^1(\R^d)$ as $\ep \to 0$. 
We replace $v$ in \refeq(11) by $v_{\ep,n}$, 
rewrite the left hand side of the resulting 
equation as      
$(G_{\th_0}\ph_n Q\nabla_{F_{\th_0}}u,
\ph_n Q\nabla_{F_{\th_0}}u) 
+ 2( \ph_n G_{\th_0} Q\nabla_{F_{\th_0} }u, 
Q(\nabla\ph_n) u)$ and arrange it as follows: 
\begin{multline} \lbeq(chu-kan)
(G_{\th_0}\ph_n Q\nabla_{F_{\th_0}}u,
\ph_n Q\nabla_{F_{\th_0}}u) 
+ (\tilde W_{\th_0} u , Q^2 \ph_n^2 u) 
= (H_1 u, Q^2 \ph_n^2 u) \\
-2(\ph_n G_{\th_0}Q\nabla_{F_{\th_0}}u, Q(\nabla \ph_n)u) 
-(G_{\th_0} Q \nabla_{F_{\th_0}} u, 
Q^{-1}\nabla(Q^2) \ph_n^2 u)
\end{multline}
By virtue of \refeq(below) the left hand side may be 
bounded from below by  
\bqn \lbeq(lhs)
(1-\th_0)\|\ph_n Q\nabla_{F_{\th_0}} u\|^2 
+ \|\ph_n Q^2 u\|^2+ (\th_0-\th)
\|\ph_n Q |B|^\frac12 u\|^2 
+ \tfrac{2C_1}{3}\|\ph_n \ax Qu \|^2 .
\eqn 
The right hand side of \refeq(chu-kan) may be bounded  
from above by  
\begin{multline}\lbeq(rhd)
\|\ph_n H_1 u\| \|\ph_n Q^2 u\|
+ 4n^{-1}\|\nabla\ph\|_{\infty}
 \|\ph_n Q \nabla_{F_{\th_0}} u\| \|Qu\| \\
+4\|\ph_n Q \nabla_{F_{\th_0}} u\| 
\|\ph_n (\nabla Q)u\| .
\end{multline}
Here we have 
$\|\ph_n (\nabla Q)u\|\leq C_Q \|\ph_n \ax Q u\|$ since 
$Q \in M(\R^d)$, and we further 
estimate \refeq(rhd) from above by 
\begin{multline}\lbeq(rhs-sch)
\tfrac1{2}\|\ph_n H_1 u\|^2+ 
\tfrac{1}{2}\|\ph_n Q^2 u\|^2 
+ 2n^{-1}\|\nabla\ph\|_{\infty}
 (\|\ph_n Q \nabla_{F_{\th_0}} u\|^2 +\|Qu\|^2)\\
+\tfrac{1-\th_0}{2}\|\ph_n Q \nabla_{F_{\th_0}} u\|^2 
+ \tfrac{8C_Q^2}{1-\th_0} \|\ph_n \ax Q u\|^2.
\end{multline}
Combining \refeq(lhs) and \refeq(rhs-sch), 
we conclude that 
\begin{align*} 
& \left(\tfrac{1-\th_0}{2}-
\tfrac{2\|\nabla \ph\|_{\infty}}{n}\right) 
\|\ph_n Q \nabla_{F_{\th_0}} u\|^2 
+ \tfrac{1}{2}\|\ph_n Q^2 u\|^2 + 
(\th_0-\th) \|\ph_n Q |B|^\frac12 u\|^2 \\
& 
\qquad+\left(\tfrac{2C_1}{3}-\tfrac{8C_Q^2}{1-\th_0}\right)
\|\ph_n \ax Q u\|^2 
\leq \tfrac1{2}\|H_1 u\|^2 + 
\tfrac{2}{n}\|\nabla\ph\|_{\infty} \|Qu\|^2.
\end{align*}
We choose $C_1>0$ larger if necessary so that 
\[
\tfrac{C_1}{6}\geq\tfrac{8C_Q^2}{1-\th_0}
\] 
and let $n\to \infty$. Then the 
monotone convergence implies that 
$Q^2 u$, $Q \nabla_{F_{\th_0}}u$, $Q|B|^\frac12 u$ 
and, a fortiori $\ax Qu $ all belong to 
$L^2(\R^d)$ and we obtain \refeq(apri-2). 
\edpf 

Since $F_{\th_0}= A + \th_0 b$ and $|b|\leq C\ax$, 
we have 
\[
(1-\th_0)\|Q\nabla_A u\|^2 \leq 
2(1-\th_0)\|Q\nabla_{F_{\th_0}} u\|^2 
+ 2C^2(1-\th_0)\th_0^2  \|\ax Q u\|^2 .
\]
Thus, assuming $2C^2<C_1$, we obtain the following 
Corollary. 

\bgcor \lbcor(previous-cor) For $\th<\th_0<1$,   
there exists $C_{\th_0}>0$ such that for 
$C_1\geq C_{\th_0}$ 
\begin{multline} \lbeq(apri-2cor)
(1-\th_0)\|Q \nabla_{A} u\|^2 + \|Q^2u\|^2 \\
+ 2(\th_0-\th)\|Q |B|^\frac12 u\|^2 
+C_1\|\ax Q u\|^2 \leq 2\|H_1 u\|^2, \quad 
u \in D(H_1).
\end{multline} 
\edcor 

Write $a_\pm=\max(0,\pm a)$ and define 
non-negative quadratic form:   
\[
q_{1+}(u)= \|\nabla_A u\|^2 
+ \|\tilde V_{1+}^\frac12 u\|^2, \quad 
D(q_{1+})=C_0^\infty(\R^d).
\]
\refth(Iwa-sa-f) implies that $q_{1+}$ is closable 
and we denote by $H_{1+}=-\nabla_A^2 + \tilde V_{1+}$
the selfadjoint operator defined by $[q_{1+}]$. 

\bglm \lblm(pre-dia) For any $\th<\th_0<1$, 
there exists $C_{\th_0}$ such that, for $C_1>C_{\th_0}$ 
we have 
\bqn \lbeq(la34)
\|\tilde V_{1-}u \|\leq (\th/\th_0)
\|H_{1+}u\|,  
\quad u \in D(H_{1+}). 
\eqn 
It follows, particular, that $D(H_1)=D(H_{1+})$. 
\edlm 
\bgpf Let $\th<\th_0<1$. 
Since $\tilde V_{1+}(x)\geq 0$, we obviously have 
\[
\th_0 |B(x)|+ \tilde V_{1+}(x) + C_\ast\ax^2  \geq 
\th_0 (1+ |B|^2+ x^4)^{1/2} 
\]
and assumption \refeq(beta-a) implies 
$Q_0(x)=\th_0^\frac12 (1+|B|^2+x^4)^{1/4}\in M(\R^d)$. 
Then, take $\th_1$ such that 
$\th_0<\th_1<1$ and repeat 
the argument of the proof of \reflm(previous) 
using $H_{1+}$, $\th_0$, $\th_1$ and $Q_0$ 
in place of $H_1$, $\th$, $\th_0$ and $Q$ 
respectively. We obtain from \refeq(apri-2) that, 
for $C_1>C_{\th_0}$, 
\bqn \lbeq(Q0xu)
\|Q_0^2(x)u\| \leq \|H_{1+}u\| , 
\quad u \in D(H_{1+}). 
\eqn 
Since 
$\tilde V_{1-} \leq \th |B(x)|$ by virtue of 
\refeq(QFM-s-imp) and 
$\th |B(x)|\leq (\th/\th_0)Q_0^2(x)$, 
\refeq(Q0xu) implies the lemma. 
\edpf   

\bgth\lbth(elli) 
There exist uniformly bounded operators 
$B_a \in \Bb(\Hg)$ for $a>0$ such that,  
for every $u \in L^2(\R^d)$, we have 
\bqn  
|(H_1 +a^2)^{-1}u(x)| 
\leq (H_{1+}+a^2)^{-1}|B_a u|(x)
\leq (-\lap + a^2)^{-1}|B_a u|(x). 
\lbeq(res-ka-in)
\eqn 
\edth 
\bgpf \reflm(pre-dia) implies that, for any $\th<\th_0<1$, 
provided that $C_1\geq C_{\th_0}$, 
\[
\|\tilde V_{1-}(H_{1+}+ a^2)^{-1}u\| 
\leq (\th/\th_0)\|u\|,  \quad u \in L^2
\]
for any $a>0$. It follows that  
\bqn \lbeq(write-po)
(H_1 +a^2)^{-1}= (H_{1+}+a^2)^{-1}B_a, \quad 
B_a= ({\bf 1} - \tilde V_{1-}(H_{1+}+a^2)^{-1})^{-1}
\eqn 
and $\|B_a\|\leq (1-(\th/\th_0))^{-1}$. We then apply 
the diamagnetic inequality (pp. 9--10 of \cite{CFKS}) 
to $H_{1+}+a^2$. The lemma follows. 
\edpf

\bgcor Provided that $C_1$ is large enough, we have 
\bqn \lbeq(elli-cor)
\|(-\lap +1)^{1/2}Q|u| \| \leq C \|H_1 u\| , \quad 
u \in D(H_1).
\eqn 
\edcor 
\bgpf \refcor(previous-cor) implies $Qu\in L^2$ and 
$\nabla_A(Qu)= Q\nabla_Au + (\nabla Q) u \in L^2$.
It follows, since $|\nabla|u||\leq |\nabla_A u|$, that 
$Q|u|\in H^1$ and  
\[
\|(-\lap +1)^{1/2}Q|u| \|^2 
= \|Qu\|^2 + \|\nabla |Qu| \|^2 
\leq \|Qu\|^2 + \|\nabla_{A}(Qu) \|^2 
\leq C\|H_1 u\|^2.   
\]
Estimate \refeq(elli-cor) follows. \edpf 

\section{Proof of \refthb(Iwap)}.  

In this and next sections we prove \refths(Iwap,Iwap-f) 
respectively. Before starting the proof, 
we briefly discuss the gauge transform which 
will play an important role in what follows.  
We define the gauge transform by 
\bqn \lbeq(gaugeG)
v(t,x)= G(t)u(t,x)= e^{-iF(t)\ax^2}u(t,x), \quad 
F(t)=\int^t_0 (C(s)+C_1)ds
\eqn 
by using a strongly 
continuous family of unitary operators 
$G(t)$, where $C_1>0$ a large constant. Then, $u(t,x)$ 
satisfies \refeq(Seq) if and only if $v(t,x)$ does  
\begin{gather} \lbeq(Seq-m)
i\pa_t v = (-\nabla_{\tilde A(t)}^2 v  + \tilde V(t,x))v, 
\\
\tilde A(t,x) = A(t,x) - 2F(t) x, \quad 
\tilde V(t,x)= V(t,x) + (C(t)+C_1)\ax^2 \lbeq(tildeV)
\end{gather} 
and, provided a dense subspace $\Si$ satisfies 
$G(t)\Si= \Si$, $\{U(t,s) \colon t,s \in \R\}$ 
is a unitary propagator for \refeq(Seq) on $\Si$ 
if and only if so is 
\bqn \lbeq(propa-tilde)
\tilde U(t,s)= G(t) U(t,s) G(s)^{-1} 
\eqn 
for \refeq(Seq-m) on $\Si$. 
If $V_1$ satisfies \refeq(bbcon-th), 
$\tilde V_1(t,x)=V_1(t,x)+ (C(t)+C_1)\ax^2$ does 
\bqn \lbeq(t-con-rev) 
|B(t,x)| + \tilde V_1 (t,x) \geq Q(x)^2+ C_1\ax^2.
\eqn  
{\it We assume in what follows 
that $C_1>0$ is taken sufficiently large} so that, 
with this $\tilde V_1(t,x)$, 
\refths(Iwa-sa,Iwa-sa-f) as well as 
\reflm(previous) and \refth(elli) are satisfied  
uniformly with respect to $t \in I$. In the proof, we 
shall first construct propagator $\tilde U(t,s)$ for 
equation \refeq(Seq-m), define $U(t,s)$ by 
\refeq(propa-tilde) and check that it satisfies the 
properties of \refth(Iwap) or \refth(Iwap-f). 

We now begin the proof of \refth(Iwap). We consider 
five operators 
\begin{gather*} 
L(t)=-\nabla_{A(t)}^2 + V(t), \ 
L_0 (t)=-\nabla_{A(t)}^2 + \tilde V(t), \  
L_1 (t)=-\nabla_{A(t)}^2 + \tilde V_1 (t), \\
\tilde L_0(t)
=-\nabla_{\tilde A(t)}^2+\tilde V(t),
\quad 
\tilde L_1(t)
=-\nabla_{\tilde A(t)}^2+\tilde V_1(t).
\end{gather*} 
These operators are all essentially selfajoint on 
$C_0^\infty(\R^d)$ and we denote their selfadjoit 
extensions by $H(t)$, $H_0(t)$, $H_1(t)$, 
$\tilde H_0(t)$ and $\tilde H_1(t)$, respectively. 

Since $V_2(t,x)$ is of Stummel 
class uniformly with respect to $t \in I$, \refth(elli) 
implies that, for any $\ep>0$, there exists $a_0$ 
such that  
\[
\|V_2(t)(H_1(t)+ a^2)^{-1}\|_{\Bb(\Hg)} \leq
\|V_2(t)(-\lap + a^2)^{-1}\|_{\Bb(\Hg)}\|B_a\|_{\Bb(\Hg)} 
<\ep , \quad a>a_0.
\]
It follows by Kato-Rellich theorem that  
\bqn \lbeq(KR-1)
H_0(t)= H_1(t) + V_2(t), \quad  
D(H_0(t))=D(H_1(t)).
\eqn 
Moreover, by choosing $C_1$ large enough we may assume 
by virtue of \refeq(apri-2),   
\[
\|u\|\leq \|H_1(t)u \|, \quad 
\|V_2(t)H_1(t)^{-1}\|\leq 1/2, \quad t \in I.
\]
Then, we have for a constant $C_0$ 
\bqn \lbeq(comparison)
C_0^{-1} \|H_1 (t) u\| 
\leq \|H_0(t) u \| \leq C_0 \|H_1 (t) u\|, \quad t \in I.
\eqn 
Since $\tilde A$ and $A$ produce the same 
magnetic field and $|\tilde A - A|\leq C\ax$, 
\refeq(KR-1) holds with $\tilde H_0(t)$ 
and $\tilde H_1(t)$ in place of $H_0(t)$ and $H_1(t)$ 
respectively and we likewise have  
\bqn \lbeq(comparison-2)
C_0^{-1} \| \tilde H_1(t) u\| \leq \| \tilde H_0(t) u\|
\leq C_0  \| \tilde H_1(t) u\| . 
\eqn

\bglm \lblm(op-equiv) 
\ben 
\item[{\rm (1)}] Domains of 
$H_0(t)$, $H_1(t)$, $\tilde H_0(t)$ and $\tilde H_1(t)$ 
satisfy  
\[
D(H_0(t))=D(H_1(t))= D(\tilde H_0(t))= D(\tilde H_1(t))
\equiv \Dg \subset D(H(t))
\]
for all $t \in I$ and $\Dg $ is independent of $t \in I$. 
\item[{\rm (2)}] There exists a constant $c>0$ such that 
\begin{gather}  
\|H_0(t) u\| \leq e^{c |t-s|}\|H_0(s)u\|, 
\quad t,s \in I, 
\lbeq(no-op-e) \\
\|(H_0(t)-H_0(s)) u\| 
\leq c |t-s| \|H_0(s)u \|, \quad t,s \in I. 
\lbeq(no-op-ee)
\end{gather}
The same holds for $\tilde H_0(t)$ replacing $H_0 (t)$. 
\item[{\rm (3)}] The gauge transform 
$G(t)= e^{-iF(t)\ax^2}$ satisfies 
$G(t) \Dg = \Dg$ and 
\bqn \lbeq(gauge-equiv)
G(t)H_0(t)  = \tilde H_0 (t)G(t), 
\quad 
G(t)H_1(t)= \tilde H_1(t)G(t). 
\eqn 
If $\ph \in \Dg$, $t\mapsto G(t)\ph$ is $\Dg$-valued 
continuous, $\Hg$-valued $C^1$ and 
\[
\pa_t G(t) \ph = -i (C(t)+C_1)\ax^2 G(t)\ph. 
\]
\een
\edlm 
\bgpf We write $C(t)$ for $C(t)+ C_1$ in the proof 
by absorbing $C_1$ into $C(t)$ for shorting formulas. 
Let $u\in C_0^\infty(\R^d)$. Then, $H_0(t)u$ is 
$\Hg$-valued differentiable almost everywhere 
with respect to $t$ and  
\bqn \lbeq(deriva)
\dot{H}_0(t)u = 2i \dot{A} (t,x) \nabla_{A(t)}u + 
i\nabla_x\cdot \dot{A} (t,x) u + \dot C(t)\ax^2 + 
\dot{V} (t,x) u. 
\eqn 
We write the right hand side in the form 
\begin{multline*}
2i \dot{A}(t,x)\cdot \nabla_{A(s,x)}u 
+2\dot{A}(t,x)\cdot\left(\int^t_s \dot{A}(r,x)dr\right)u 
+ (i\nabla_x\cdot \dot{A} (t,x)+\dot C(t)\ax^2) u \\
+ \dot{V} (t,x) u 
= I_1(t,s) u + I_2 (t,s) u + I_3 (t) u + I_4(t)u. 
\end{multline*}
Since $|\dot A(t,x)|\leq C Q(x)$, \refeq(apri-2cor) 
implies 
\[
\|I_1(t,s) u \| \leq 2\||\dot A(t,x)||\nabla_{A(s)}u|\|
\leq C \|Q \nabla_{A(s)}u\| \leq C \|H_1(s)u\|.
\] 
Denote by $M(t,x)$ any of 
$\nabla_x (\dot{A} (t,x)^2)$, 
$\dot{A} (t,x)^2$, $\nabla_x \cdot \dot{A} (t,x)$ and 
$\dot C(t)\ax^2$. Then, $|M(t,x)| \leq CQ(x)^2$ 
and \refeq(apri-2cor) implies  
$\|M(t)H_1(s)^{-1}u\|\leq C_1 \|u\|$  
uniformly with respect to $t,s \in I$. Thus, 
\[
\|I_2(t,s)u\|+ \|I_3(t)u\|\leq C \|H_1(s) u\|,  
\quad t,s \in I.
\]  
Write $\dot V(t,x)= W_0(t,x)+ W_1(t,x)+ W_2(t,x)$ 
as in \refth(Iwap), then 
$\|W_0(t)u\|\leq C\|Q^2 u\|\leq C\|H_1(s)u\|$ 
for any $t,s\in I$ as above; 
\[
\|W_1(t)u\| 
\leq \|Q^{-1}W_1(t)(-\lap +1)^{-\frac12} \|_{\Bb(\Hg)}
\|(-\lap+1)^{\frac12}Q |u|\|\leq C \|H_1(s)u\|
\]
by virtue of \refeq(elli-cor); and \refth(elli) implies  
\[
\|W_2(t)H_1(s)^{-1} u\|
\leq C \|W_2(t)(-\lap+1)^{-1}|B_1 u|\| 
\leq C \|B_1 u\| \leq C\|u\|. 
\]
Thus, $\|I_4(t,s)u\|\leq C \|H_1(s)u\|$ 
and combining these estimates, we obtain  
\bqn \lbeq(ope-eq0)
\|\dot{H}_0(t)u\| \leq C \|H_1(s)u\|  
\leq C\|H_0(s)u\| , \quad t,s \in I. 
\eqn 
It follows by integration that 
\bqn \lbeq(ope-eq)
\|(H_0(t)-H_0(s))u\|\leq c |t-s| \|H_0(s)u\|, 
\quad u \in C_0^\infty(\R^d) .
\eqn 
Since $C_0^\infty(\R^d)$ is a core of 
$H_0(s)$, \refeq(ope-eq) extends to 
$u \in D(H_0(s))$. It follows that 
$D(H_0(s)) \subset D(H_0(t))$ and 
by symmetry $D(H_0(s))=D(H_0(t))$ 
for any $t,s \in I$ and, consequently, 
\refeq(no-op-ee) for $H_0(t)$
is satisfied. \refeq(no-op-ee) clearly 
implies \refeq(no-op-e). 
Changing $A(t)$ by $\tilde A(t)$ will not change $B(t,x)$ 
and the argument above yields the 
same results for $\tilde H_0(t)$ and $\tilde H_1(t)$.  
This proves statement (2). 

Let $u \in D(H_0(t))$. 
Then, $\ax^2 u \in \Hg$ by virtue of 
\refeq(apri-2cor) and 
\bqn \lbeq(maxH)
H (t) u = H_0(t)u -C(t)\ax^2 u \in \Hg. 
\eqn 
Since $D(H(t))=\{u \in \Hg \colon H(t) u \in L^2\}$, 
\refeq(maxH) implies $u \in D(H(t))$ and 
$D(H_0(t)) \subset D(H(t))$. 

We next prove $D(H_1(t))=D(\tilde H_1(t))$, which will 
then prove statement (1). Define for $\th \in [0,1]$ 
\[
H_1 (t, \th) = -\nabla_{A(t,\th)}^2 + 
\tilde V_1(t,x),
\quad A(t,\th,x) = A(t,x) - 2\th F(t) x, 
\]
so that $H_1 (t,0)=H_1(t)$ and 
$H_1 (t,1)=\tilde H_1(t)$. Since $A(t,\th,x)$ 
and $A(t,x)$ generate the same magnetic field $B(t,x)$ 
and $|2\th F(t) x|\leq C \ax$, 
results of previous sections apply to $H_1(t,\th)$.  
We have  
\[
\pa_{\th} H_1(t,\th)u 
= -i4F(t)x\nabla_{A(t)}u +8\th F(t)^2x^2 u
-2diF(t)u 
\] 
and \refeq(apri-2cor) implies 
$\|\pa_{\th} H_1(t,\th)u\|\leq C \|H_1(t)u\|$ for 
$0\leq \th \leq 1$. Thus,  
\[
\|(H_1 (t,\th) - H_1 (t,\s))u\|
\leq C |\th-\s|\|H_1(t,\s)u\|, 
\quad u \in C_0^\infty(\R^d), 
\] 
and we obtain the desired result 
$D(H_1(t))=D(\tilde H_1(t))$ as previously.  

It is clear that $G(t)$ is an isomorphism of 
$C_0^\infty(\R^d)$ and 
$G(t)H_0(t)\ph = \tilde H_0(t)G(t)\ph$ 
for $\ph \in C_0^\infty(\R^d)$.
Since $C_0^\infty(\R^d)$ is a core of $H_0(t)$, 
it follows that 
$G(t)D(H_0(t)) \subset D(\tilde H_0(t))$. This 
clearly holds for $G(-t)=G(t)^{-1}$ as well and we obtain  
$G(t)\Dg = \Dg $ and $G(t)H_0(t)= \tilde H_0(t)G(t)$.
This argument likewise applies to the pair 
$H_1(t)$ and $\tilde H_1(t)$ 
and we obtain \refeq(gauge-equiv). 
The last statement is obvious since 
$\Dg \subset D(\ax^2)$. 
This completes the proof. \edpf

\noindent
{\bf Proof of \refthb(Iwap)}. \reflm(op-equiv) 
yields statement (a) of the theorem. 
It also implies that graph norms of any two of 
$\{H_0(t), \tilde H_0(s)\colon t,s \in I\}$ 
are equivalent to each other. We equip $\Dg$ 
with the graph norm of $H_0(t_0)$ as in the theorem. 
Then, it is obvious that $\Dg \subset \Hg$ continuously 
and densely, $\Dg =D(\tilde H_0(t))$ for every 
$t\in I$ and that 
$I \ni t \mapsto \tilde H_0(t)\in \Bb(\Dg, \Hg)$ 
is norm continuous by virtue of \refeq(no-op-ee) 
for $\tilde H_0(t)$. We wish to apply \refth(se) 
to the triplet $(\Xg, \Yg, A(t))$ by setting 
$\Xg=\Hg$, $\Yg=\Dg$ and $A(t)=\tilde H_0(t)$. 
For this we need check conditions (1) and 
(2) of \refth(se) are satisfied. 

For $t \in I$, we define $\Yg_t=\Dg$ but with 
the graph norm of $\tilde H_0(t)$ and $\Xg_t = \Hg$. 
Then, the norm of $\Yg_t$ is 
equivalent to that of $\Dg$ and 
\refeq(no-op-e) for $\tilde H_0(t)$ 
implies condition \refeq(xy-equiv). 
It follows from \refth(LS) that $\tilde H_0(t)$ is 
selfadjoint in $\Xg_t=\Hg$. Hence the part 
of $\tilde H_0(t)$ in $\Yg_t(=D(\tilde H_0(t)))$ is 
automatically selfadjoint with domain 
$D(\tilde H_0(t)^2)$. 
Thus, the conditions are satisfied. 

It follows that there uniquely exists a family of 
operators $\{\tilde U(t,s) \colon s, t\in I\}$ which 
satisfies properties of \refth(se) for 
$(\Hg, \Dg, \tilde H_0(t))$. 
Moreover, $\tilde U(t,s)$ is a unitary operator of $\Hg$. 
Indeed, if we set 
$u(t) = \tilde U(t,s)\ph$ for $\ph \in \Yg$,  
$i\pa_t \|u(t)\|^2 = (\tilde H_0(t) u(t), u(t))- 
(u(t), \tilde H_0(t)u(t))=0 $
since $ \tilde H_0(t)$ is selfadjoint. 
Hence $\tilde U(t,s)$ is an isometry of 
$\Hg$ and, since $\tilde U(t,s)\Dg=\Dg$, it 
is unitary. We define 
\[
U(t,s) = G(t)^{-1} \tilde U(t,s) G(s). 
\]
Then, $U(t,s)$ is a strongly continuous family 
of unitary operators on $\Hg$; \reflm(op-equiv) (3) 
implies that $U(t,s)\in \Bb(\Dg)$; if 
$\ph \in \Dg$, $U(t,s)\ph$ is $\Dg$-valued 
continuous, $\Hg$-valued $C^1$ and that $U(t,s)\ph$ 
satisfies the first of Eqns. \refeq(dieq-1):
\[
i \pa_t U(t,s)\ph = 
G(t)^{-1}(-C(t)\ax^2 + \tilde H_0(t)) 
\tilde U(t,s) G(s)\ph  
=H(t)U(t,s)\ph.  
\] 
We may similarly prove that $U(t,s)\ph$ 
satisfies the other of \refeq(dieq-1). 

For proving the uniqueness of $U(t,s)$ 
we have only to notice the following: If $U(t,s)$ 
satisfies properties of the theorem, then 
$\tilde U(t,s) = G(t)U(t,s)G(s)^{-1}$ does those  
for $\tilde H_0(t)$ and such 
$\tilde U(t,s)$ is unique by virtue of \refth(se). 

When $\ph \in \Dg$, \refeq(dieq-1) shows that  
$u(t,x)=U(t,s)\ph(x)$ satisfies 
\refeq(Seq) in the sense of distributions. 
Then, the standard approximation argument shows that 
the same holds for $\ph \in \Hg$ as well 
and $U(t,s)$ is unitary propagator on $\Hg$ for 
\refeq(Seq). We omit the details. 
The proof is completed. \qed 

\section{Proof of \refthb(Iwap-f)} 

For the constant $\th$ in \refeq(QFM-Iwap) 
we take and fix $\th_0$ such that 
$\th<\th_0<1$ and take the constant $C_1>0$ large enough 
so that results of \refsecs(self,diam) are 
satisfied, uniformly with respect to $t \in I$, 
for $q_0(t)$ of \refeq(q0q) and 
\[
q_1(t)(u,v)=(\nabla_{A(t)}u, \nabla_{A(t)}v)  
+ (\tilde V_1(t)u, v) , \quad u, v \in C_0^\infty(\R^d),
\]
in place of $q_0$ and $q_1$ respectively. 
In addition to $q_0(t)$ and $q_1(t)$, we define  
\begin{gather} 
\lbeq(q0-def)
\tilde q_0(t)(u,v) = 
(\nabla_{\tilde A(t)} u, \nabla_{\tilde A(t)} v) 
+ (\tilde V u , v), \quad u, v \in C_0^\infty(\R^d),\\
\lbeq(q1-def)
\tilde q_1(t)(u,v) = 
(\nabla_{\tilde A(t)} u, \nabla_{\tilde A(t)} v) 
+ (\tilde V_1 u , v), \quad u, v \in C_0^\infty(\R^d),
\end{gather}
where $\tilde A(t,x) = A(t,x) - 2F(t)x$. 
Since $\tilde A(t,x)$ and 
$A(t,x)$ generate same magnetic field and they differ 
only by $2F(t)x$, results of 
\refsecs(self,diam) likewise apply to 
$\tilde q_0(t)$ and $\tilde q_1(t)$ 
uniformly for $t\in I$. 
In particular, since $V_2$ is of Kato class uniformly with 
respect to $t \in I$, $\tilde q_1(t)$ is uniformly 
positive definite and 
\bqn  
C^{-1}\tilde q_1(t)(u)\leq \tilde q_0(t)(u) 
\leq C \tilde q_1(t)(u), \quad u \in C_0^\infty(\R^d)
\eqn
for a $t$-independent constant $C >0$. 
Thus, $D([q_0(t)])=D([q_1(t)])$ and 
$D([\tilde q_0(t)])=D([\tilde q_1(t)])$. 
We denote by $H_0(t)$, $H_1(t)$, $\tilde H_0(t)$ and 
$\tilde H_1(t)$ selfadjoint operators 
defined respectively by $[q_0(t)]$, $[q_1(t)]$, 
$[\tilde q_0(t)]$ and $[\tilde q_1(t)]$. 
As in the previous section, we write $C(t)$ for $C(t)+C_1$ 
absorbing $C_1$ into $C(t)$. 

\bglm \lblm(q-equiv) \ben 
\item[{\rm (1)}] Domains of  
$[q_0(t)]$, $[q_1(t)]$, $[\tilde q_0(t)]$ and 
$[\tilde q_1(t)]$ satisfy 
\[
D([q_0(t)])=D([q_1(t)])=D([\tilde q_0(t)])
=[\tilde q_1(t)]=\Yg \subset D(L_Q^{\frac12})
\]
and are independent of $t\in I$.   
\item[{\rm (2)}] There exists a constant $c>0$ such that 
\bqn \lbeq(y-equiv)
[\tilde q_0(t)](u) \leq e^{c|t-s|}[\tilde q_0(s)](u), 
\quad 
\quad u \in \Yg, \quad t, s \in I. 
\eqn 
\item[{\rm (3)}] The gauge transform $G(t)$ maps 
$\Yg$ onto $\Yg$ and 
\bqn \lbeq(q-equiv10)
[\tilde q_0(t)](G(t)u)=[q_0(t)](u), \ \ 
[\tilde q_1(t)](G(t)u)= [q_1(t)](u), \ \ u \in \Yg. 
\eqn 
\een
\edlm 
\bgpf By virtue of \refeq(below) corresponding to 
$\tilde A(t,x)$ and $\tilde V(t,x)$, we have 
\bqn \lbeq(eq25)
\|Qu\|^2 + \|\nabla_{\tilde A(t)} u\|^2 
\leq C\tilde q_0(t)(u), 
\quad u \in C_0^\infty(\R^d), \quad t\in I .
\eqn 
Hence, $\|\dot{\tilde A} (t)u\|^2 \leq C \|Q u\|^2 
\leq C\tilde q_0(s)(u)$ 
for any $t,s\in I$ and by integration 
\bqn \lbeq(aeast)
\|(\tilde A(t)- \tilde A(s))u\|\leq 
C|t-s|\tilde q_0(s)(u)^{\frac12}.  
\eqn 
Likewise, using,  in addition to 
\refeq(eq25), assumption \refeq(previous) 
and obvious identity $\||\dot{\tilde V} (r)|^{1/2}u\|
=\||\dot{\tilde V} (r)|^{1/2}|u|\|$, we obtain that 
\[
\||\dot{\tilde V} (r)|^{1/2}u\|^2  
\leq C (\|\nabla|u|\|^2+ \|Qu\|^2)
\leq C (\|\nabla_{\tilde A(s)}u\|^2+ \|Qu\|^2) 
\leq C \tilde q_0(s)(u).
\]
Applying this to 
$\tilde V(t,x)-\tilde V(s,x)
=\int_s^t \dot{\tilde V}(r,x)dr$, 
we have  
\bqn \lbeq(veast)
|((\tilde V(t)-\tilde  V(s))u, v)|
\leq C |t-s|\tilde q_0(s)(u)^\frac12 
\tilde q_0(s)(v)^\frac12. 
\eqn 
Write $\tilde q_0 (t)(u,v) -\tilde q_0(s)(u,v)$ 
for $u,v \in C_0^\infty(\R^d)$ in the form 
\begin{multline*}
(\nabla_{\tilde A(s)}u, i(\tilde A(s)-\tilde A(t))v)  
+ (i(\tilde A(s)-\tilde A(t))u, \nabla_{\tilde A(s)}v) \\ 
+ ((\tilde A(t)-\tilde A(s))u, 
(\tilde A(t)-\tilde A(s))v)+ 
((\tilde V(t)-\tilde V(s))u, v). 
\end{multline*}
We estimate each term separately by using \refeq(eq25), 
\refeq(aeast) and \refeq(veast). 
We obtain for $|t-s|\leq 1$ that 
\bqn   
|\tilde q_0(t)(u,v) -\tilde q_0(s)(u,v)|   
\leq C|t-s|\tilde q_0(s)(u)^\frac12 
\tilde q_0(s)(v)^\frac12. 
\lbeq(qh3-est) 
\eqn  
It follows that 
$D([\tilde q_0(t)])= D([\tilde q_0(s)])$ as in the 
proof of \reflm(op-equiv), all estimate above extend 
to $u,v$ in $D([\tilde q_0(t)])=D([\tilde q_0(s)])$ 
and 
\bqn \lbeq(q0-tscom)
[\tilde q_0(t)](u) \leq (1+ C|t-s|)[\tilde q_0(s)](u) 
\leq e^{C|t-s|}[\tilde q_0(s)](u).
\eqn 
Argument above applies to $q_0(t)$ as well and we have 
\refeq(eq25) for $u\in D([q_0(t)])$; 
$D([q_0(t)])= D([q_0(s)])$ for $t,s \in I$; and    
estimate \refeq(q0-tscom) holds for 
$[q_0(t)]$ and $[q_0(s)]$. Moreover, we have 
$D([q_1(t)])= D([\tilde q_1(t)])$ 
by virtue of characterization formula \refeq(q0q-dom) 
of domains of the forms. Since 
$\|L_Q^{\frac12}u\|^2 \leq 
C (\|Qu\|^2 + \|\nabla_{\tilde A(t)} u\|^2)$ 
for $u \in C_0^\infty(\R^d)$, we also have 
$D([\tilde q_0(t)]) \subset D(L_Q^\frac12)$ 
from \refeq(eq25). Statements (1) and (2) follow. 

Both 
$\|\nabla_{\tilde A(t)} G(t)u\| = \|\nabla_{A(t)}u\|$ 
and $(V(t)G(t)u, G(t)u)= (V(t)u,u)$ are obvious for 
$u \in C_0^\infty(\R^d)$. Since the latter space is a core 
of the forms $[q_0(t)]$ and $[\tilde q_0(t)]$, we see that  
$D([\tilde q_0(t)])= G(t) D([q_0(t)])$, 
$G(t)$ maps $\Yg$ onto $\Yg$, and that 
$[\tilde q_0(t)](G(t)u)= [q_0(t)](u)$ for $u \in \Yg$.  
The corresponding relation for 
$[q_1(t)]$ and $[\tilde q_1(t)]$ may be proved similarly. 
\edpf 

Before proceeding to the proof \refth(Iwap-f), 
we recall the following general fact: 
If $H$ is a 
positive selfadjoint operator in a Hilbert space $\Hg$, 
$\Hg_1 \subset \Hg \subset \Hg_{-1}$ 
is the scale of Hilbert spaces associated with 
$H$, viz. $\Hg_1=D(H^{1/2})$  
and $\Hg_{-1}= \Hg_1^\ast$ with $\Hg^\ast$ 
being identified with $\Hg$, then:
\ben 
\item[{\rm (i)}] $\Hg_{-1}$ is the completion of 
$\Hg$ by the norm $\|H^{-1/2}u\|$. 
\item[{\rm (ii)}] $H$ has a natural extension 
$H_{-}$ to $\Hg_{-1}$ and $H_{-}$ is selfadjoint 
in $\Hg_{-1}$ with domain $D(H^{1/2})$. 
\item[{\rm (iii)}] The part $H_{+}$ of 
$H_{-}$ in $\Hg_{1}$ is again 
selfadjoint with domain $D(H^{3/2})$. 
\een
These should be obvious if, by using spectral 
representation theorem, we represent $H$ as a 
multiplication operator by a positive function on 
$L^2(M, d\m)$, $(M, d\m)$ being a suitable measure space.  

\noindent 
{\bf Proof of \refthb(Iwap-f)}. We equip $\Yg$ with 
the inner product $q_0(u,v)$ and let $\Xg$ be 
its dual space as in the theorem. It is obvious 
that $\Yg\subset \Xg$ densely and continuously. 
\reflm(q-equiv) yields 
statement (a) except for the fact that 
$H(t)\in \Bb(\Yg, \Xg)$ and it is norm continuous. 
To prove the latter fact, we first show that 
the multiplication by $\ax^2$ is bounded from 
$\Yg$ to $\Xg$ by using \refeq(eq25) for 
$q_0(t)$: 
\begin{multline}
\|\ax^2 u\|_{\Xg} =
\sup_{v\in \Yg, \|v\|_{\Yg}=1}|(\ax^2 u, v)| 
\leq C \sup_{v\in \Yg, \|v\|_{\Yg}=1}\|Q u\| \|Q v\| \\
\leq C \sup_{v\in \Yg, \|v\|_{\Yg}=1}
[q_0(t_0)](u)^{\frac12}
[q_0(t_0)](v)^{\frac12}
= C\|u\|_{\Yg}.   \lbeq(mult-11)
\end{multline}
Then, we estimate for $u,v \in C_0^\infty(\R^d)$ via 
\refeq(q-equiv10) for $[q_0(t)]$ as follows:
\[
|(H(t)u, v)| \leq |q_0(t)(u,v)| + |(C(t)\ax^2 u,v)|
\leq C(e^{2c|t-t_0|}+ C(t))\|u\|_{\Yg}\|v\|_{\Yg}.
\]
and $\|H(t)u\|_{\Xg} \leq C \|u\|_{\Yg}$. This extends 
to $u \in \Yg$ since $C_0^\infty(\R^d)$ is dense in $\Yg$. 
Thus, $H(t) \in \Bb(\Yg,\Xg)$. We have   
\begin{multline*}
((H(t)-H(s))u, v)=((H_0(t)-H_0(s))u, v)
-((C(t)-C(s))\ax^2 u,v)\\
= ((q_0(t)-q_0(s))u,v)
- ((C(t)-C(s))\ax^2 u,v), \quad u, v \in \Yg. 
\end{multline*}
Thus, \refeq(qh3-est) for $q_0(t)$ and \refeq(mult-11) 
imply 
$\|H(t)-H(s)\|_{\Bb(\Yg,\Xg)} \leq C(|t-s|+ |C(t)-C(s)|)$ 
and statement (a) follows. 

We define $\Yg_t$ to be $\Yg$ with new inner 
product $(u,v)_{\Yg_t}=[\tilde q_0(t)](u,v)$ 
and $\Xg_t$ to be the dual space of 
$\Yg_t$ with respect to the inner product of $\Hg$. 
Then, $\Xg_t \subset \Hg \subset \Yg_t $ 
is the scale of Hilbert space associated with 
positive selfadjoint operator $\tilde H_0(t)$.  
Then, by virtue of the property (i), 
$\Xg_t$ is independent of $t$ as a set and is 
equal to $\Xg$ since $\Yg_t=\Yg$ is independent of 
$t$ as a set with equivalent Hilbert space structures. 
Properties (ii) and (iii) produce selfadjoint     
operators $\tilde H_0(t)_{-}$ and $\tilde H_0(t)_{+}$ 
in $\Xg_t$ and $\Yg_t$ respectively. It is evident 
that $\tilde H_0(t)_{-}$ is a closed operator in $\Xg$ 
(with respect to the original norm) and 
$\tilde H_0(t)_{+}$ is its part in $\Yg$. 
We now want to apply  \refth(se) to triplet 
$(\Xg, \Yg, \tilde H_0(t)_{-})$. 

We check conditions of \refth(se) for 
$(\Xg, \Yg, \tilde H_0(t)_{-})$. Norm  
$\|u\|_{\Yg_t}$ is equivalent with the original 
one of $\Yg$ by virtue of the closed graph theorem.  
Estimate \refeq(y-equiv) implies that 
$\{\|u\|_{\Yg_t}\colon t \in I\}$ satisfies 
condition \refeq(xy-equiv) of \refth(se) for 
$\Yg_t$ and likewise for $\Xg_t$ by duality. 
From \refeq(qh3-est) we have   
\bqn \lbeq(tlah0)
|\la (\tilde H_0(t)_{-}-\tilde H_0(s)_{-})u, v \ra| \leq 
c|t-s|\tilde q_0(s)(u)^\frac12 \tilde q_0(s)(v)^\frac12,
\eqn 
where $\la \cdot, \cdot\ra$ on the left is the coupling 
between $\Xg$ and $\Yg$. This implies that 
\bqn 
\|(\tilde H_0(t)_{-}-\tilde H_0(s)_{-})u\|_{\Xg_s} 
\leq c|t-s|\|u\|_{\Yg_s} 
\eqn 
and we see that 
$I \ni t \to \tilde H_0(t)_{-}\in \Bb(\Yg, \Xg)$ 
is norm continuous. 

Thus, there uniquely exists a family 
of operators $\{\tilde U(t,s)\colon t,s \in I\}$ 
which satisfies the properties of \refth(se) 
for $(\Xg, \Yg, \tilde H_0(t)_{-})$. We define 
\[
U(t,s) = G(t)^{-1}\tilde U(t,s) G(s).
\] 
We know that $G(t)$ maps $\Yg$ onto $\Yg$ by virtue of 
\reflm(q-equiv) and, \refeq(mult-11) implies 
that, for $u \in \Yg$, $I \ni t \mapsto G(t)u\in \Xg$ is 
continuously differentiable.  Then, it is easy to 
check that $U(t,s)$ is satisfies all properties 
of statement (b) except that $U(t,s)$ is a strongly 
continuous family of unitary operators in $\Hg$, 
which we now show.  
Define $u(t)=U(t,s)\ph$ for $\ph \in \Yg$. 
Then, with $\la \cdot,\cdot \ra$ being the coupling 
of $\Xg$ and $\Yg$, we have 
\begin{align*} 
\pa_t (u(t), u(t))_{L^2}& = 2\Re \la -i H(t)u(t), 
u(t) \ra  \\
& = 2\Re\{-iq_0(t)(u(t), u(t))+ 
iC(t)\la \ax^2 u(t), u(t)\ra \}=0 .  
\end{align*} 
It follows that $\|u(t)\|=\|\ph\|$ 
and, since $\Yg$ is dense in $\Hg$, 
we conclude 
$U(t,s)\Hg \subset \Hg$ and $\|U(t,s)\ph\|=\|\ph\|$ 
for all $\ph \in \Hg$. Then, 
$U(t,s)$ must be unitary since $U(t,s)U(s,t)\ph=\ph$. 
If $\ph \in \Yg$, $(t,s) \mapsto U(t,s)\ph\in \Hg$ is  
continuous in $\Hg$. Hence $U(t,s)$ is strongly continuous 
in $\Bb(\Hg)$ by the unitarity. 
The uniqueness of $U(t,s)$ of \refth(Iwap-f) follows from the uniqueness result of \refth(se)  
by tracing back the argument above. 
\qed

\bibliographystyle{amsalpha}

\end{document}